\documentclass{ws-ijmpe}
\usepackage[super,compress]{cite}

\newcommand{\be}{\begin{equation}}
\newcommand{\ee}{\end{equation}}

\newcommand{\nn}{\nonumber}
\newcommand{\bea}{\begin{eqnarray}}
\newcommand{\eea}{\end{eqnarray}} 

\newcommand{\la}{\langle}
\newcommand{\ra}{\rangle}

\newcommand{\Z}{\mathbb{Z}}
\newcommand{\R}{{\kern+.25em\sf{R}\kern-.78em\sf{I} \kern+.78em\kern-.25em}}
\newcommand{\RR}{{\kern+.25em\sf{R}\kern-.6em\sf{I} \kern+.6em\kern-.25em}}
\newcommand{\N}{{\kern+.25em\sf{N}\kern-.78em\sf{I} \kern+.78em\kern-.25em}}
\newcommand{\C}{\mathbb{C}}

\newcommand{\ri}{{\rm i}}

\newcommand{\vp}{\varphi}

\newcommand{\gtapprox}{\raisebox{-0.5ex}{$\,\stackrel{>}{\scriptstyle\sim}\,$}}
\newcommand{\ltapprox}{\raisebox{-0.5ex}{$\,\stackrel{<}{\scriptstyle\sim}\,$}}

\begin{document}

\markboth{Wolfgang Bietenholz}{Hadron Physics from Lattice QCD}

\catchline{}{}{}{}{}
%%%%%%%%%%%%%%%%%%%%%%%%%%%%%%%%%%%%%%%%%%%%%%%%%%%%%%%%%%%%%%%%%%%%

\title{Hadron Physics from Lattice QCD}

\author{Wolfgang Bietenholz}

\address{Instituto de Ciencias Nucleares \\
Universidad Nacional Aut\'{o}noma de M\'{e}xico \\
A.P. 70-543, C.P. 04510 Ciudad de M\'{e}xico, Mexico\\
wolbi@nucleares.unam.mx}

\maketitle

\begin{history}
\received{Day Month Year}
\revised{Day Month Year}
%\accepted{Day Month Year}
%\comby{(xxxxxxxxxx)}
\end{history}

\begin{abstract}

We sketch the basic ideas of the lattice regularization in
Quantum Field Theory, the corresponding Monte Carlo simulations,
and applications to Quantum Chromodynamics (QCD). This approach enables
the numerical measurement of observables at the non-perturbative
level. We comment on selected results,
with a focus on hadron masses and the link to Chiral 
Perturbation Theory.
At last we address two outstanding issues:
topological freezing and the sign problem.

\end{abstract}


\ccode{PACS numbers: 11.15.Ha, 12.38.Gc}

%\tableofcontents

\section{Introduction}

Since the 1970s, QCD is generally accepted as the fundamental
theory that underlies nuclear physics. It is formulated in terms
of {\em quark} and {\em gluon fields,} but what our detectors
actually observe are {\em hadrons.} In contrast to quarks and gluons,
hadrons are composite particles, color singlets with an extremely 
complicated internal structure. One distinguishes {\em baryons} 
(which contain three valence quarks) and {\em mesons} (with a 
valence quark--anti-quark pair). 

Of course, hadrons do not just {\em consist} of these valence
quarks, although some quantum numbers are obtained correctly
if we sum up the valence quark contributions. If we consider the
mass, however, we see that this simplification fails: 
in particular the lightest quark flavors, $u$ and $d$, have masses
in the range of $\approx 2 \dots 6 \, {\rm MeV}$, which are provided 
by the Higgs mechanism. Hence the valence quark content of a 
nucleon ($p \sim (uud)$, $n \sim (udd)$) contributes only 
${\cal O}(1) \, \%$ to the nucleon mass, 
$M_{p,n} \approx 939 \, {\rm MeV}$.\footnote{Even more extreme is 
the case of a {\em glueball,} which does not contain any valence quarks.}

We conclude that most of the masses of the macroscopic objects
around us do {\em not} emerge from the
Higgs mechanism. The dominant contribution
is encoded in
a dense tangle of gluons, along with
virtual quark--anti-quark pairs (sea quarks). The situation
is similar for other (light) hadrons. Deriving their masses from 
{\em first principles of QCD} has been a major challenge since the 1970s.
Perturbation theory is inappropriate for this purpose, since the
strong coupling ``constant'' is large at low energy (Section 3
specifies the reference scale).
A non-perturbative derivation, and therefore a stringent test of 
QCD at low energy, has been a primary goal of {\em lattice QCD}
from its beginning.
The basic concepts of this approach were elaborated in the 1970s and 
1980s, and a breakthrough in its applications has been achieved in 
recent years.

Section 2 summarizes the ideas of the lattice
regularization in Quantum Field Theory, and of the
corresponding Monte Carlo simulations --- for further
details we refer to text books\cite{books}.
In Section 3 we comment on results for hadron masses, 
and the relation  between lattice QCD and Chiral Perturbation Theory.
Section 4 adds concluding remarks, and two appendices refer to
selected outstanding issues.

\section{The lattice regularization of Euclidean Quantum Field Theory}

Our framework is the functional integral formulation of
Quantum Field Theory in Euclidean space (in natural units,
$\hbar = c =1$). This formulation provides
a link to Statistical Mechanics, from where we adopt terms like the
{\em partition function}
\be
Z = \la 0 | 0 \ra = \int {\cal D} \Phi \ e^{- S [ \Phi ]} \ ,
\ee
where $\Phi (x)$ represents some field, $S$ is the Euclidean
action (a functional of the field configuration $[ \Phi ]$), and
${\cal D} \Phi$ is the functional measure for the integration 
over all configurations.
Most relevant observables take the form of $n$-point functions,
{\it i.e.}\ expectation values of the form\footnote{Variants of this
form appear in QCD: {\it e.g.}\ for static mesons,
the r\^{o}le of $\Phi(x_k)$ is taken by two factors of the form 
$\sum_{\vec x_k} \bar \psi_{2} (\vec x_k, t_k) \, \Gamma \, \psi_{1} 
(\vec x_k, t_k )$, in the slices at times $t_k$, $k=1,2$, where 
$\psi_{1}$, $\psi_{2}$ are quark fields for two flavors, and $\Gamma$ is an 
element of the Clifford algebra, cf.\ Subsection 3.1.\label{fermdense}}
\be
\la 0 | T \hat \Phi (x_1) \dots \hat \Phi (x_n) | 0 \ra =
\frac{1}{Z} \int {\cal D} \Phi \ \Phi (x_1) \dots \Phi (x_n) \ 
e^{- S [ \Phi ]} \ .
\ee
The left-hand-side refers to the canonical formalism
(with an operator-valued field $\hat \Phi$ and the time ordering
operator $T$), while the right-hand-side expresses the same
quantity in the functional integral form, in analogy to thermal 
expectation values in Statistical Mechanics, 
and $x_{k}=(\vec x_{k}, t_{k})$.
We adopt the interpretation as a statistical system and interpret
\be  \label{prob}
p [ \Phi ] = \frac{1}{Z} \ e^{- S [ \Phi ]}
\ee
as the {\em probability} of the configuration $[ \Phi ]$.\footnote{At
this point, we assume $S$ to be real positive for any configuration.
In fact, this holds in many situations of interest.
If this is not the case, we face a serious problem, known as the
``sign problem''; this an outstanding issue, to be 
addressed in Appendix B.\label{signprob}}

\subsection{Lattice regularization}

Calculations in Quantum Field Theory require an UV regularization,
which preserves the symmetries, or allows them to be restored in
the final UV limit. 
The lattice regularization is a simple but powerful scheme:
it reduces the Euclidean space (or space-time)
to discrete sites $x$, which carry the matter 
field variables (gauge fields are defined on the links, 
see Subsection 2.4). The most popular structure is
a simple hyper-cubic lattice, with a spacing that we denote as $a$,
which means (in $d$ dimensions) $x/a \in \Z^{d}$.
One often uses {\em lattice units} by setting $a=1$.

On the regularized level, Poincar\'{e} symmetry is reduced to a
discrete form, but the restoration of this global symmetry
in the continuum limit is conceptually on safe 
ground.\footnote{Additional terms may enter the regularized system, 
but they are {\em irrelevant} in the sense of the Renormalization Group, 
so they vanish in the continuum limit. What matters most
is that {\em local} symmetries are conserved on the lattice, 
{\it i.e.}\ gauge invariance holds, see Subsection 2.4.}
A continuum matter field $\Phi (x)$ is reduced to $\Phi_{x}$,
so it is only defined on the lattice sites $x$.
Thus the momenta are confined to the (first) Brillouin zone, 
$p \in ( -\pi/a,\pi/a]^{d}$, {\it i.e.}\ we impose an UV cutoff 
$\pi /a$ on each component $|p_{\mu}|$. The 
(initially mysterious) functional integral simplifies to 
the well-defined form
\be
\int {\cal D} \Phi \ \to \ \int \prod_{x} d \Phi_{x} \ ,
\ee
where the integrals run over all (allowed) field values at 
each lattice site $x$. Now the functional measure has an
explicit meaning, but typically the number of integrals is
far too large to be computed directly. Instead it can be
handled by {\em importance sampling,} to be described next.

\subsection{Lattice simulations}

The idea of Monte Carlo simulations of a lattice regularized
Quantum Field Theory is to generate a large set of lattice
configurations $[\Phi ]$, which are random distributed according
to the probability given in 
eq.\ (\ref{prob}), $p[\Phi ] \propto \exp (- S[\Phi ])$. 
If this is achieved, these ``golden configurations'' specify values
of the observables of interest (such as an $n$-point function);
averaging over these values constitutes a {\em numerical measurement.}
As in experiments, the result is correct up to statistical
and systematic errors:

\begin{itemize}

\item {\em Statistical errors:} the set of ``golden'' 
({\it i.e.}\ useful) configurations
generated in a simulation is necessarily finite, but the total set
(which should actually be integrated over) is usually 
infinite.\footnote{An exception is {\it e.g.}\ the Ising model,
with field values $\sigma_{x} \in \{1,-1\}$,
in some lattice volume $V$. However,
the total number of configurations, $2^{V}$, tends to be huge:
for instance, on a simple $32^{3}$ lattice it amounts to $\approx
10^{9864}$, which can hardly be summed over. Even here one
would resort to importance sampling, which provides
quite precise results based on relatively few configurations.}
One has to invest an amount of CPU time, which is sufficient for 
generating and analyzing a number of configurations that leads to 
small statistical errors, such that the result is conclusive.
We add that one can often perform
multiple measurements in one configuration, {\it e.g.}\ of the
correlation function (or 2-point function) $\la \Phi_{x} \Phi_{y} \ra$
over a fixed distance $|x-y|$, so even a modest number of 
configurations may provide considerable statistics.

\item {\em Systematic errors:} Simulations must be carried out at finite
lattice spacing, $a >0$, in a finite volume $V < \infty$, but in
Quantum Field Theory we are generally interested in the {\em continuum limit}
$a \to 0$ (UV limit) and the {\em infinite volume limit} $V \to \infty$
(IR limit).\footnote{Exceptions occur in solid state physics ($a$ 
could be a physical spacing in a crystal), or in the $\epsilon$-regime
of QCD (where one studies light mesons in a small physical volume,
cf.\ Subsection 3.3).} 

The natural scale of the system, which decides what we consider
as ``large'' or ``small'', and how far we are from these limits,
is set by the {\em correlation length} $\xi$. It characterizes 
the exponential decay of the connected correlation function
$\la \Phi_{x} \Phi_{y} \ra_{c} = 
\la \Phi_{x} \Phi_{y} \ra - \la \Phi_{x} \ra \la \Phi_{y} \ra$,
over a long distance $|x-y| \gg a$, 
\be  \label{decay}
\la \Phi_{x} \Phi_{y} \ra_{c}
\propto \left\{ \begin{array}{cccc}
\exp (- |x-y|/\xi) &&& V = \infty \ \\
\cosh ([\, |x-y| - La/2]/\xi ) &&& V = (aL)^{d}. \, \end{array} \right. 
\ee
In case of a finite volume $V$, we assume periodic boundary
conditions (for bosonic fields), {\it i.e.}\ a torus,
such that discrete translation invariance holds.
We require $a \ll \xi \ll L$, and the final limits 
$\xi /a , \ L/\xi \to \infty$ lead to a {\em critical point,} 
with a phase transition of second (or higher) order.
So we are dealing with {\em critical phenomena,} as Kenneth 
Wilson --- the leading pioneer of lattice field theory --- and 
others pointed out.\cite{WilKog}

One extrapolates to these limits based on simulation 
results at various $a$ and $V$, and theoretical knowledge about
the form of the leading artifacts. Finite size effects are often
exponentially suppressed in $L/\xi$. The dominant source of 
systematic errors tends to be the lattice artifacts, 
{\it e.g.}\ in purely bosonic theories they set in 
at ${\cal O}((a/\xi)^{2})$. The uncertainty in these extrapolations
--- and sometimes further ones, cf.\ Section 3 --- can be 
estimated by established methods, see {\it e.g.}\ Ref.\ \refcite{numrep}.

\end{itemize}

As a great virtue, the lattice approach is fully 
{\em non-perturbative.} At no point the action is split into
a free and an interaction part, as it is done in perturbation theory,
which relies on an expansion of the form
$$
\exp (-S) = \exp(-(S_{\rm free} + S_{\rm int})) \approx 
\exp(-S_{\rm free}) \ [1 - S_{\rm int} +  S_{\rm int}^{2}/2 \dots ] \ .
$$
In contrast, here the entire action $S$ is left in the exponent,
where it belongs. It fixes the probabilities of the configurations,
cf.\ eq.\ (\ref{prob}), so one works
directly at {\em finite} interaction strength. This allows
us to capture even settings of strong coupling, like QCD
at low energy: this case is relevant {\it e.g.}\ for the nucleon 
masses, and generally for hadron physics under ordinary conditions, 
where perturbation theory is inapplicable ($S_{\rm int}$ is too 
large to be treated as an expansion quantity).

\subsection{Monte Carlo methods}

The Monte Carlo algorithms, which are used in this context, 
generate a long sequence of configurations,
\be
[ \Phi ] \to [ \Phi' ] \to [ \Phi'' ]  \to [ \Phi''' ] \dots \ .
\ee
Each new configuration is generated based on the previous one, 
without considering the earlier history; this is a {\em Markov chain}. 

As a first example, we describe the simple but robust
{\em Metropolis algorithm}\footnote{Its application to the
anharmonic oscillator is described very pedagogically
in Ref.\ \refcite{CreutzFreedman}.}:
an update step starts with some {\em suggestion} for a
new configuration, which is often a small random modification of 
the previous one, where only the field variable in one site or link
changes. The decision whether or not this suggestion is accepted
has to obey the condition of {\em Detailed Balance,}
\be  \label{balance}
\frac{p [ \Phi_{1} \to \Phi_{2} ]}{p [ \Phi_{2} \to \Phi_{1} ]} =
\frac{p[ \Phi_{2} ]}{p[ \Phi_{1} ]} = e^{- \Delta S[\Phi_{1},\Phi_{2}]} \ , \quad
\Delta S[\Phi_{1},\Phi_{2}] = S[ \Phi_{2} ] - S[ \Phi_{1} ] \ , 
\ee
where $p [ \Phi_{i} \to \Phi_{j} ]$ is the acceptance probability of a
transition from configuration $[ \Phi_{i} ]$ to  $[ \Phi_{j} ]$.
Moreover, an algorithm has to be {\em ergodic:} starting from any
configuration, any other (allowed) configuration must be accessible
within a finite number of update steps (the probability for
attaining it must be non-zero).

If the algorithm respects these conditions, and we perform
a large number of update steps (from any initial configuration),
then we will obtain configurations with the required probability
distribution (\ref{prob}). They should be independent of each other,
so the configurations to be used in the numerical measurements
--- we called them ``golden configurations'' ---
have to be separated by a significant number of update steps;
we have to suppress their {\em auto-correlation.}

The Metropolis algorithm implements Detailed Balance with
the prescription
\be
p_{\rm Metropolis} [ \Phi_1 \to \Phi_2 ] = \left\{ \begin{array}{ccccc}
1 &&&& {\rm if} \ \Delta S[\Phi_{1},\Phi_{2}] \leq 0 \\
\exp(- \Delta S[\Phi_{1},\Phi_{2}] ) &&&& {\rm otherwise.} \end{array} \right.
\ee
This algorithm is simple to implement and widely applicable, 
though there are often more efficient alternatives. 

For instance the {\em heatbath} algorithm considers 
a small part of the configuration,
say $\Phi_{x}$, computes the probability for its possible values
in the background of the fixed rest of the configuration (the 
``heatbath''), and selects $\Phi_{x}'$ randomly with this probability.
Here the previous value $\Phi_{x}$ doesn't matter, and
no accept/reject step is necessary.
The implementation is more difficult than the Metropolis algorithm, 
it is only feasible in certain models, but if it works it is superior.
In particular, the heatbath algorithm is standard in simulations 
of pure gauge theories (without matter fields).

{\em Molecular Dynamics} prepares a new configuration
based on the Hamiltonian equations of motion, such that it is
--- in principle --- automatically accepted, due to Liouville's 
Theorem. In practice this in not exact, since one follows the 
Hamiltonian trajectory in discrete jumps. Therefore a Metropolis 
accept/reject step is added nevertheless, but one obtains a 
high acceptance rate.

{\em Cluster algorithms}\cite{clusteralgo} construct --- in a 
stochastic manner --- ``clusters'' of field variables to be updated
collectively, again without needing an accept/reject decision (the
corresponding probabilities are anticipated in the cluster 
formation). This can be highly efficient, since one proceeds in
the space of all configurations ``in large leaps'', rapidly
suppressing auto-correlations. Moreover, cluster algorithms often
enable the use of an {\em Improved Estimator,} which allows for
the statistical inclusion of numerous configurations, without 
actually generating them. It is extremely powerful when applied 
to certain spin models\cite{clusteralgo}, 
and fermionic models\cite{clusterfermi}, 
but no efficient cluster algorithm is known 
for gauge theories --- unfortunately.

The standard algorithm for QCD with (dynamical) quarks is 
called {\em Hybrid Monte Carlo} (HMC) \cite{HMC}. It 
extends the action by a
momentum field, with a Gaussian probability distribution. Then one 
applies Molecular Dynamics to follow the Hamiltonian trajectory
({\em Langevin algorithm}\cite{ParisiWu}).
The field and momentum are updated in small alternating steps,
structured according to the {\em Leap frog} method 
(or something similar) to closely follow the Hamiltonian
trajectory over some length; 
deviations are corrected at the end by a Metropolis step.
Now the momentum field is refreshed with a new random Gaussian, 
and the next trajectory begins. 
The application to QCD requires --- in addition to the gauge field 
--- an auxiliary ``pseudo-fermion field'' to avoid the computation of 
the fermion determinant, see Subsection 2.5.\\

\begin{figure}[h!]
\includegraphics[width=4.68cm,angle=270]{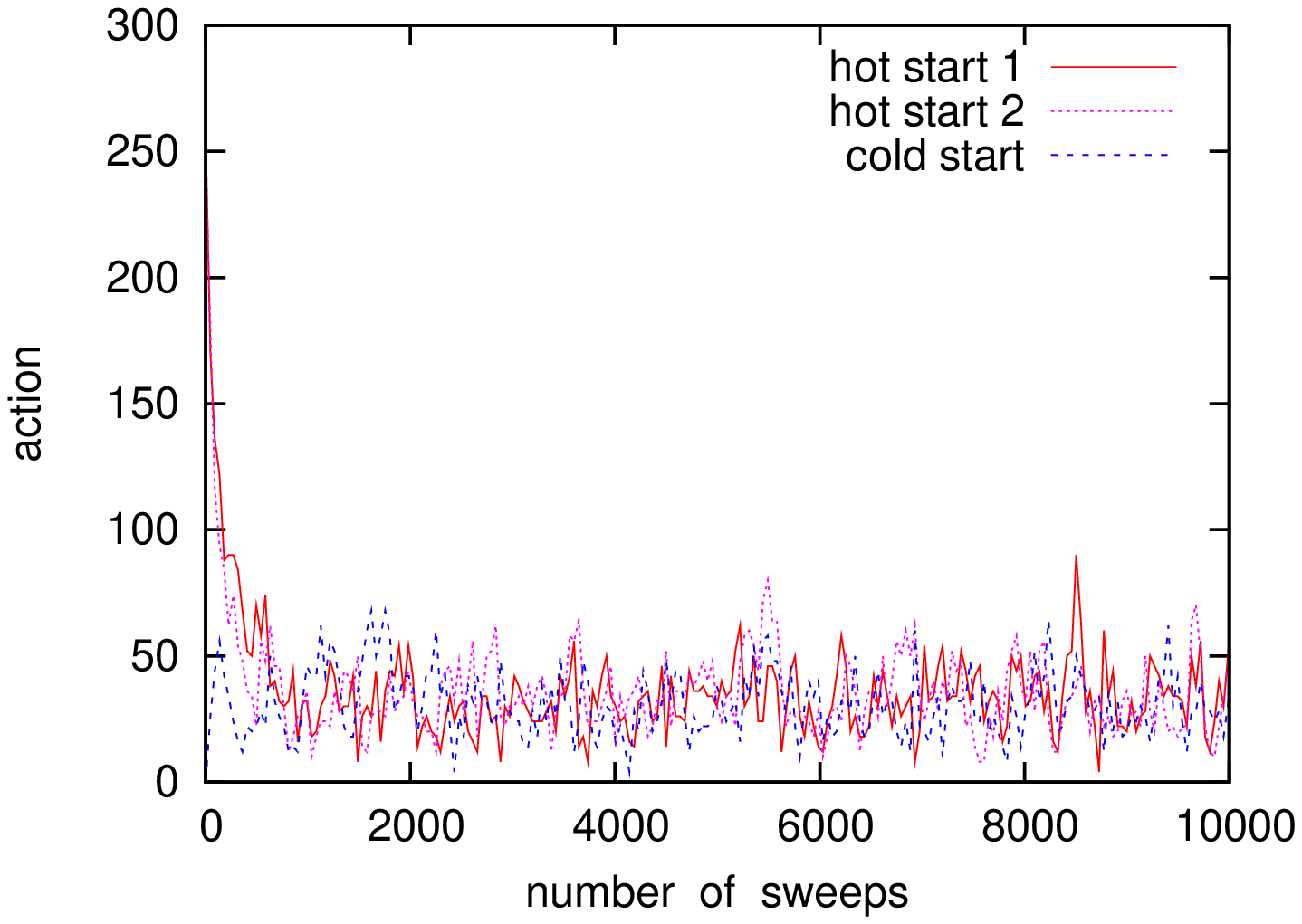}
\hspace*{-3mm}
\includegraphics[width=4.28cm,angle=270]{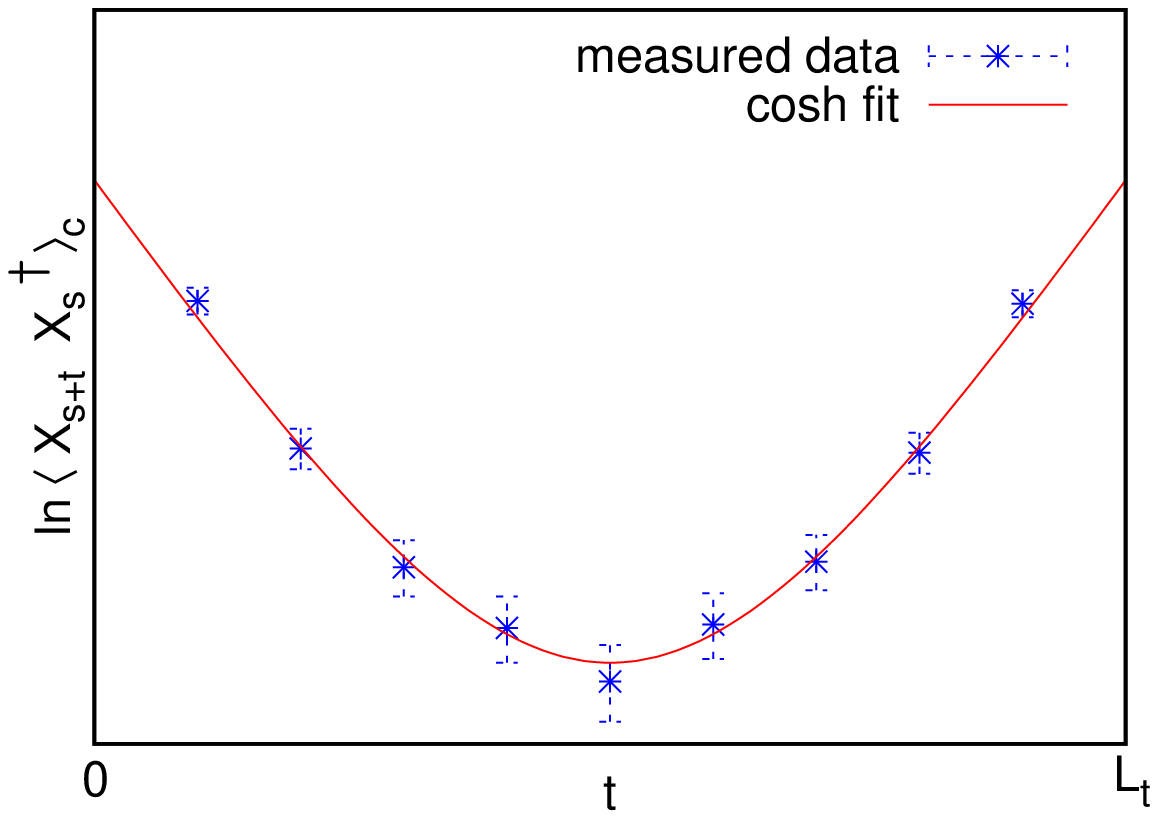}
\caption{On the left: examples for Monte Carlo histories, {\it i.e.}\
the evolution of the Euclidean action $S$ depending on the number of
update steps in units of ``sweeps'' (one update for each site of 
the lattice volume). We show the beginning of three histories, with 
two ``hot starts'' and one ``cold start'', which soon fluctuate 
around the same stable plateau (at $\la S \ra \approx 33$). 
On the right: possible results for a connected correlation function,
$\la X_{s+t} X_{s}^{\dagger} \ra_{c}$ (in $d>1$, $X_{s}$ could be a sum over 
a time slice, $X_{s} = \sum_{\vec x} X_{\vec x ,s}$, which corresponds 
to momentum $\vec p = \vec 0$, cf.\ footnote \ref{fermdense}),
and a fit to the expected cosh-function of eq.\ (\ref{decay}), 
for periodic boundary conditions over the extent $L_{t}$ in Euclidean 
time. The physical interpretation of $L_{t}$  is an inverse 
temperature, see Appendix B.}
\label{thermo}
\vspace*{-4mm}
\end{figure}

{\it A priori,} we do not know what would be a particularly suitable
configuration to start a simulation, so we start anywhere: 
a {\em cold start} begins with a hand-made trivial configuration, 
it could be uniform. The opposite is a {\em hot start} from a ``wild''
or ``rough'' random configuration, {\it e.g.}\ in case of the 
Ising model
on chooses $\sigma_{x} = \pm 1$ with equal probability at each site $x$.
Such wildly fluctuating configurations have a high Euclidean action 
(in spin models one would call it ``energy''), and the algorithm
will easily find ways to decrease it. Thus
the action drops rapidly, until it reaches an equilibrium,
as illustrated in Fig.\ \ref{thermo} (left) for
some model with minimal action $S [\Phi _{\rm classical}] = 0$.

At this point, the configurations are quite smooth. The
Boltzmann factor $\exp(-S)$ favors a further decrease of $S$, 
but this requires specific, even smoother configurations. There are
more random update suggestions which would increase $S$, but these
are less likely to be accepted, due to detailed balance (\ref{balance}). 
The equilibrium between a Boltzmann
effect (pulling $S$ down) and an entropy effect (pushing $S$ up) causes
fluctuations around a stable mean value. This regime is independent
of the initial configuration (for a cold start it is attained from 
below), and {\em here} the numerical measurement can be performed.
The first part of the Monte Carlo history, the {\em thermalization},
is discarded --- it is just needed for first taking the Markov chain
to the right regime, the {\em thermal equilibrium.}

Now one still has to assure that the thermal equilibrium 
configurations that one uses are (essentially) independent of each 
other. To this end, {\it i.e.}\ in order to avoid {\em auto-correlation}
effects, the measurements have to be separated by a sufficient 
number of update steps.\footnote{\label{autocor}The absence of 
significant auto-correlation can be verified statistically.\cite{numrep} 
The required separation depends on the algorithm, and also on the
observable under investigation (as a functional of the configuration).} 
In the example illustrated in Fig.\ \ref{thermo} (left), one 
should first wait for a few thousand Monte Carlo sweeps for the 
thermalization. If we subsequently measure the expectation value 
of the action itself, $\la S \ra$, the separation should be a few 
hundred sweeps --- for subtle observables it has to be longer, 
cf.\ footnote \ref{autocor}.

Once we have generated
a sizable set of ``golden configurations'' (well thermalized
and decorrelated), we perform the numerical measurement. Fig.\ 
\ref{thermo} (right) shows, as an example, possible results
for the connected correlation function
of some quantity $X$ (a product of fields, often denoted as an ``operator'')
over different separations $t$ in Euclidean time. 
A fit to the expected $\cosh$ function, cf.\ eq.\ 
(\ref{decay}), yields a result for the correlation length, which 
represents the inverse first energy gap, $\xi = 1/(E_{1}-E_{0})$.
In quantum field theory, this first excitation energy above the
vacuum is the mass of the (lightest) particle which corresponds to
the quantity $X$.

\subsection{Lattice gauge theory}

Let us consider a free, complex scalar field, $\Phi_{x} \in \C$,
with the lattice action
\bea
S[ \Phi ] &=& \frac{a^{d-2}}{2} \sum_{x,y} \Phi_{x}^{*} M_{xy} \Phi_{y} \ , \nn \\
M_{xy} &=& \sum_{\mu =1}^{d} (-\delta_{x + a \hat \mu,y} 
-\delta_{x - a \hat \mu,y} + 2 \delta_{x,y}) + (ma)^{2} \delta_{x,y} \ ,
\label{comscalact}
\eea
where $\hat \mu$ is the unit vector in $\mu$-direction.
This is the most obvious lattice discretization of the
continuum action 
$S[ \Phi ] = \int d^{d}x \, \Phi (x)^{*} \, 
[ - \partial^{2} + m^{2}] \, \Phi (x)$. 

\begin{figure}[h!]
\vspace*{-3mm}
\begin{center}
\includegraphics[width=6cm,angle=0]{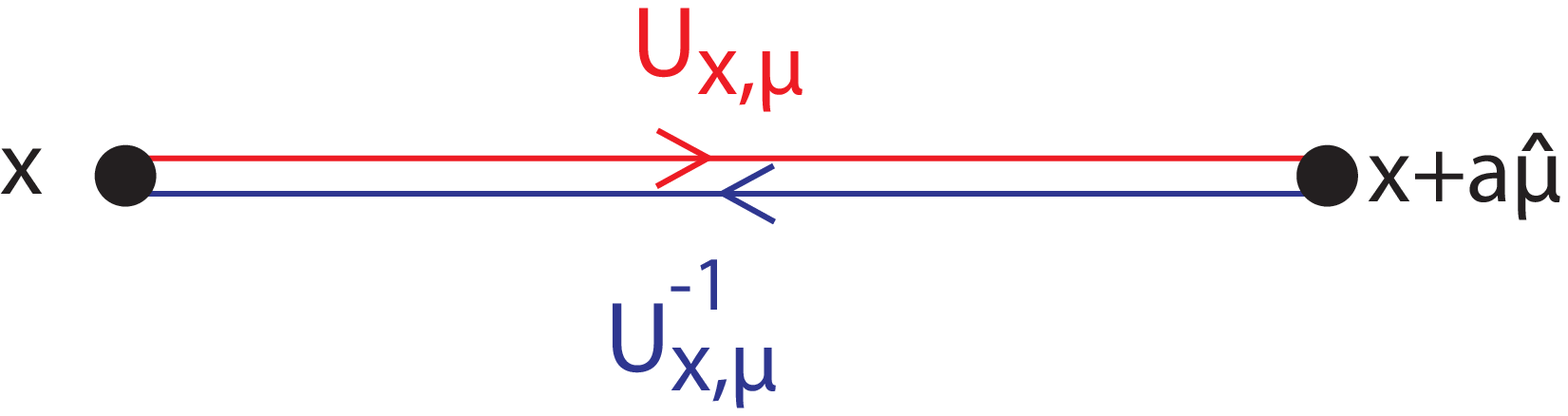}
\hspace*{5mm}
\includegraphics[width=5cm,angle=0]{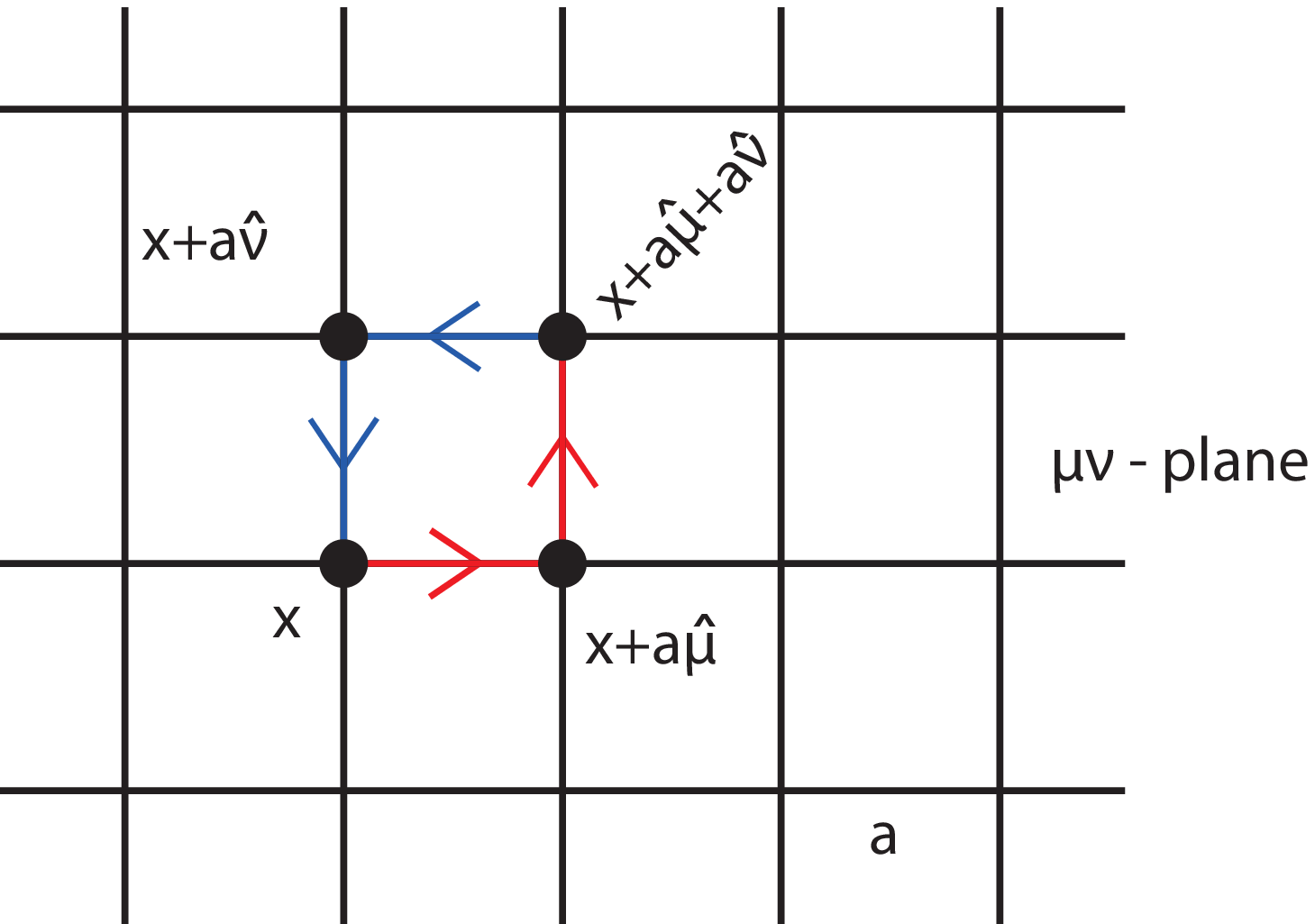}
\vspace*{-3mm}
\end{center}
\caption{On the left: an compact link variable 
$U_{x,\mu} \in \{ {\rm gauge~group} \}$, as part of the
lattice gauge field. On the right: illustration of
a plaquette variable, as given in eq.\ (\ref{plaq}).}
\label{gaugefig}
\vspace*{-4mm}
\end{figure}

We now follow the procedure, which is standard in the continuum,
but we do so on the lattice. We promote the global U(1) symmetry 
$\Phi_{x} \to e^{\ri g \vp} \Phi_{x}$ to a {\em local} symmetry,
$\vp \to \vp_{x}$. To this end,
we replace the $\delta$-links in $M$, $\delta_{x \pm a \hat \mu,y}$, 
by link variables in U(1), which transform such that gauge symmetry 
holds. To be explicit, we substitute in eq.\ (\ref{comscalact}) 
\be
\Phi_{x}^{*} \Phi_{x+a\hat \mu} \longrightarrow
\Phi_{x}^{*} U_{x,\mu} \Phi_{x+a\hat \mu} \ , \
\Phi_{x}^{*} \Phi_{x - a\hat \mu} \longrightarrow
\Phi_{x}^{*} U_{x-a \hat \mu ,\mu}^{*} \Phi_{x-a\hat \mu} \ , 
\quad U_{x,\mu} \in {\rm U}(1) \ ,  \nn
\ee
as illustrated in Fig.\ \ref{gaugefig} (left), such that a gauge 
transformation
\be
\Phi_{x} \to e^{\ri g \vp_{x}} \Phi_{x} \ , \quad
U_{x,\mu} \to e^{\ri g \vp_{x}} U_{x,\mu} e^{-\ri g \vp_{x+ a \hat \mu}}  
\ee
leaves the lattice action invariant. So we have implemented a 
discrete covariant derivative, which endows gauge invariance 
{\em at the regularized level.}
Generically, such a {\em compact link variable} is an element of 
the {\em gauge group} (not of its algebra!), 
{\it e.g.}\ $U_{x,\mu} \in {\rm SU}(N_c)$;
then the link from $x$ in $-\mu$-direction takes the form 
$U_{x-a\hat \mu, \mu}^{\dagger}$.

To construct the lattice gauge action, we first build the 
{\em plaquette variable}\,\footnote{Most literature writes the
link factors in inverse order, which leads
to the same lattice gauge action, but we consider the order 
given here more obvious.}
\be  \label{plaq} 
U_{x,\mu \nu} : = U_{x,\nu}^{-1} U_{x + a\nu , \mu}^{-1}
U_{x + a\mu , \nu} U_{x, \mu} \in \{ \mbox{\,gauge~group\,} \} \ ,
\ee
which represents a minimal lattice Wilson loop in the $\mu \nu$-plane, 
as shown in Fig.\ \ref{gaugefig} (right). Therefore 
$U_{x,\mu \nu}$ is gauge invariant as well, and a suitable ingredient
for the lattice gauge action (Wilson's standard formulation)
\bea
S_{\rm gauge} [U] &=&
\frac{2N_c}{g^{2}} \sum_{x, \, \mu < \nu} \Big( 1 - 
\frac{1}{2 N_{c}} {\rm Tr} [U_{x,\mu \nu} + U_{x,\mu \nu}^{\dagger}] \Big) \nn \\
&=& \frac{2N_c}{g^{2}}  \sum_{x, \, \mu < \nu} \Big( 
1- \frac{1}{N_{c}} {\rm Re} \, {\rm Tr} \, U_{x,\mu \nu} \Big)
\ , \ \ {\rm for~}U_{x,\mu}, \, U_{x,\mu \nu} \in {\rm SU}(N_{c}). \quad
\label{Sgauge}
\eea
In contrast to continuum gauge theory, no gauge fixing is
needed, thanks to the use of compact link variables. This 
is a great relief, since it avoids nightmares related to Faddeev-Popov
ghost fields. Moreover, compact link variables simplify the update
procedure in the Monte Carlo simulation.\footnote{While gauge
fixing is not {\em a priori} necessary on the lattice (in contrast
to the continuum), it is sometimes done nevertheless, for
practical reasons, like the suppression of statistical noise 
and the measurement of specific quantities, in particular quark
and gluon propagators.}

In order to confirm that the continuum limit
takes the expected form, we introduce non-compact link variables
$A_{x,\mu}$ (in the algebra), and the corresponding continuum
field $A_{\mu}(x)$. For instance in the simple case of 4d U(1)
gauge theory we obtain
\bea
U_{x,\mu} &=& \exp \Big( -\ri g \int_{x}^{x+ a \hat \mu} ds \, A_{\mu}(s) \Big)
= e^{-\ri g a A_{x,\mu}} \ , \
U_{x,\mu \nu} = e^{- \ri a^{2} g F_{x,\mu \nu}} \ , \nn \\
F_{x,\mu \nu} & = & \frac{(A_{x+a \hat \mu, \nu} - A_{x, \nu}) -
(A_{x+a \hat \nu, \mu} - A_{x, \mu})}{a}
= F_{\mu \nu}(x) + {\cal O}(a) \ , \nn \\
S_{\rm gauge}[U] &=& \frac{1}{g^{2}} \sum_{x, \, \mu < \nu}
(1 - {\rm Re} \, U_{x,\mu \nu} ) = 
\frac{1}{4} \int d^{4}x \, F_{\mu \nu}(x) F_{\mu \nu}(x) + {\cal O}(a^{2}) \ .  
\eea

\subsection{Fermion fields}

We treat $\Psi (x)$ and $\bar \Psi (x)$ as independent fermion fields,
and assume an action of bilinear form,\footnote{This form captures 
virtually all models of interest. Even models with a 4-Fermi term,
$(\bar \Psi \Psi)^{2}$, such as the 2d Gross--Neveu model and the 4d 
Nambu--Jona-Lasinio model, can be written in 
this form by means of an auxiliary scalar field.} with the partition 
function
\be
Z = \int {\cal D} \bar \Psi \, {\cal D} \Psi \,
\exp (- \bar \Psi M \Psi ) =
\int \prod_{k} d \bar \psi_{k} \, d \psi_{k} \,
\exp \Big( - \sum_{ij} \bar \psi_{i} M_{ij} \psi_{j} \Big) \ .
\ee
The components, given by the indices $i,j,k$, run over everything,
{\it i.e.}\ over all 
lattice sites, and on each site over the internal degrees of freedom, 
such as the spinor index and possible indices for $N_{f}$ flavors and
$N_{c}$ colors.

For each spinor, the fermion matrix $M$ contains a discrete, Euclidean
Dirac operator $D$. In contrast to the scalar field action
in Subsection 2.4, its formulation is not straightforward,
not even for free fermions. The na\"{\i}ve discretization of the linear
derivative, $\partial_{\mu} \Psi (x) \to \frac{1}{2a} (\Psi_{x + a \hat \mu}
- \Psi_{x - a \hat \mu})$, fails. It is plagued by the notorious
{\em fermion doubling} problem: its free propagator in momentum space,
\be
\tilde D_{\mbox {na\"{\i}ve}}(p)^{-1} = 
\Big[ \ri \gamma_{\mu} \frac{1}{a} \sin (a p_{\mu}) + m \Big]^{-1}
= \frac{- \ri \gamma_{\mu} \frac{1}{a} \sin (a p_{\mu}) + m}
{\frac{1}{a^{2}} \sum_{\mu} \sin^{2} (a p_{\mu}) + m^{2}} \ ,
\ee
has --- in the chiral limit $m\to 0$ --- $2^{d}$ poles, instead
of one (in the Brillouin zone).

A number of valid formulations are known,
such as the {\em Wilson fermion}:\cite{Wilfer}
$D_{\mbox {na\"{\i}ve}}$ is supplemented by a discrete Laplacian, 
which moves the masses of the doublers to the cutoff scale,
\bea
D_{\rm Wilson} &=& \frac{1}{2} \sum_{\mu =1}^{d}
\Big[ \gamma_{\mu} (\nabla_{\mu} + \nabla_{\mu}^{*})
- a \nabla_{\mu}^{*} \nabla_{\mu} \Big] + m \ , \nn \\
{\rm free~fermion:} && 
\nabla_{\mu} \psi_{x} = (\psi_{x + a \hat \mu} - \psi_{x})/a \ , 
\quad \nabla_{\mu}^{*} \psi_{x} = (\psi_{x} - \psi_{x - a \hat \mu})/a \ ,
\label{DWilson}
\eea
where $\nabla_{\mu}$ ($\nabla_{\mu}^{*}$) is a forward (backward)
lattice derivative.
Unfortunately the Wilson term explicitly breaks the chiral symmetry 
of massless fermions, $\{ D_{\rm Wilson}, \gamma_{5} \} \neq 0$,
which entails problems like an additive mass renormalization
(see below).

The other standard formulation is known as 
{\em staggered fermions:}\cite{KoSu} 
a transformation replaces the $\gamma$-matrices by $x$-dependent
sign factors. In $d=4$ this reduces the number of doublers by a
factor of 4 (the number of spinor components), 
and the remaining $16/4 = 4$ species are used
as a kind of degenerate flavors (``tastes''). 
Here a ${\rm U}(1)$ subgroup of the chiral symmetry persists.
However, $N_{f}=4$ light quark flavors is not what we need in QCD.
Formally this formulations can be generalized to one (and therefore to any) 
number of flavors by taking the forth root of the fermion determinant
(see below), but locality is questionable for such ``rooted 
staggered fermions''.\footnote{The continuum limit is conceptually 
safe ({\it i.e.}\ the formulation is guaranteed to be
in the right universality class) if the couplings decay at least
exponentially in the distance $|x-y|$ between $\bar \Psi_{x}$ and
$\Psi_{y}$. This is the meaning of ``locality'' in (most of) the
lattice literature.}

Since the end of the 20th century we also have formulations which preserve 
the full chiral flavor symmetry, ${\rm SU}(N_{f}) \otimes {\rm SU}(N_{f})$,
cf.\ Subsection 3.3, 
in a lattice modified form, which turns into the standard form in the 
continuum limit\cite{ML}.\footnote{Here we refer to ``vector theories'',
like QCD, where the left-handed and right-handed fermions have the
same gauge couplings. They differ in ``chiral gauge theories'',
like the electroweak sector of the Standard Model, where
a non-perturbative regularization is even more difficult; for 
achievements in that respect, see {\it e.g.}\ Ref.\ \refcite{ML2}.}
The condition for this property is the {\em Ginsparg-Wilson 
Relation,}\cite{GiWi} which reads (in its simplest form)
\be
\{ D , \gamma_{5} \} = a D \gamma_{5} D \ .
\ee 
Various types of solutions are known, such as Domain Wall 
Fermions\cite{Kaplan}, (classically) perfect fermions\cite{perfect} 
and overlap fermions\cite{overlap}, but they are all rather tedious 
to simulate. For reviews we refer to Refs.\ \refcite{revGWR}.

In the presence of a gauge field $U_{x,\mu}$, all these formulations 
of the lattice Dirac operator contain covariant lattice derivatives,
\be
\nabla_{\mu} \psi_{x} = \frac{1}{a} (U_{x, \mu} \psi_{x + a \hat \mu} - \psi_{x}) 
\ , \quad \nabla_{\mu}^{*} \psi_{x} = 
\frac{1}{a} (\psi_{x} - U_{x - a \hat \mu , \mu}^{-1}\psi_{x - a \hat \mu}) \ ,
\ee 
hence also the fermion action is gauge invariant
at the regularized level.
Regarding the Wilson term, $-\frac{a}{2} \nabla_{\mu}^{*} \nabla_{\mu}$,
the link couplings are
suppressed, whereas the on-site term $\propto \bar \psi_{x} \psi_{x}$
remains invariant, which leads to the aforementioned additive 
mass renormalization (along with further nuisance, like 
${\cal O}(a)$ scaling artifacts).
Hence a negative bare mass $m$ has to be
tuned in order to get close to chirality; in QCD this means attaining
light pions. This tuning --- to search for criticality --- is not needed 
in the cases of staggered fermions or Ginsparg-Wilson 
fermions, thanks to the (remnant) chiral symmetry on the lattice.\\

In the canonical formalism of Quantum Field Theory, Pauli's principle 
--- as part of the Spin-Statistics Theorem --- is implemented by the 
anti-commutation behavior of the fermionic creation and annihilation 
operators. They do not occur in the functional integral formulation, 
so one implements the fermionic anti-commutativity by the type of fields:
the components $\bar \psi_{i}$, $\psi_{j}$ are given by {\em Grassmann
variables,} {\it i.e.}\ elements $\eta_{1},\eta_{2},\eta_{3} \dots$ 
of a Grassmann algebra, with the rules
\be
\{ \eta_{i}, \eta_{j} \} = 0 \ , \quad
\frac{\partial}{\partial \eta_{i}} \eta_{j} = \delta_{ij}
= \int d \eta_{i} \ \eta_{j} \ ,
\ee
where the integral in the last 
term does not have any bounds. It is defined such that it obeys
translation invariance (see {\it e.g.}\ Ref.\ \refcite{PS}).

Application of these rules leads to the celebrated formulae
\be  \label{fermdet}
\int {\cal D} \bar \Psi \, {\cal D} \Psi \,
\exp (- \bar \Psi M \Psi ) = {\rm det} \, M \ , \quad
\la \bar \psi_{i} \, \psi_{j} \ra = - (M^{-1})_{ij} \ ,
\ee
where ${\rm det}\, M$ is the {\em fermion determinant.}
In light of these results, we see that computers do not 
need to handle Grassmann variables, they ``just''
deal with the (complex) fermion matrix $M$. 
However, in typical QCD simulations $\bar\Psi$, $\Psi$
have millions of components, hence capturing ${\rm det}\, M$ 
and $M^{-1}$ is a formidable task, the {\em bottleneck} 
in lattice QCD simulations.

In 4d lattice gauge theories, a (frequent) computation of 
${\rm det} \,M$ is hardly feasible. The HMC algorithm\cite{HMC}
circumvents this problem as follows: first we note that all
usual lattice Dirac operators $D$ are {\em $\gamma_{5}$-Hermitian,}
\be  \label{g5H}
D^{\dagger} = \gamma_{5} D \gamma_{5}
\quad \Rightarrow \quad H := \gamma_{5} D = H^{\dagger} \ .
\ee
For two degenerate flavors, the fermion determinant is expressed
by introducing an auxiliary multi-scalar field $\Phi \in \C^{N}$,
which is denoted as a ``pseudo-fermion field'',
\be
{\rm det} \, D[U]^{2} = {\rm det} \, H[U]^{2} 
= \int {\cal D} \Phi^{\dagger} {\cal D} \Phi \ 
e^{- G^{\dagger}[U,\Phi ] \, \cdot \, G[U,\Phi ]} \ , \ 
G[U,\Phi ] := H[U]^{-1} \Phi \ .
\ee
Updating the pseudo-fermion field is not necessary thanks to 
the Gaussian structure in $G$; this random distribution can
be generated directly, as it is also done for the
momentum field in the HMC algorithm; both are refreshed before
beginning a new Molecular Dynamics trajectory.
(There are quite efficient methods to iteratively invert a huge but
sparse, positively definite matrix, like $D^{\dagger}D = H^{2}$.)

Once the fermion fields are integrated out, the configuration
is given in terms of the gauge link variables, $[U]$. If we
update just one link, the change of the gauge action, 
$\Delta S_{\rm gauge}$ in a notation analogous to eq.\ (\ref{balance}),
can be computed locally, since it only affects $2(d-1)$ plaquette
variables. 
In the presence of fermions, however, ${\rm det}\, M$ alters $S$ in 
a complicated manner. Now numerous degrees of freedom are coupled, 
and the HMC algorithm is appropriate, since it modifies the entire 
configuration at once. 
Still, the inclusion of quarks makes the generation of QCD 
configurations much more tedious: the computational effort 
increases by (about two) orders of magnitude.

\section{Lattice QCD and the hadron spectrum}

\subsection{Set-up}

As discussed in Section 2, a lattice QCD configuration $[U]$
is given by a set of link variables $U_{x,\mu} \in {\rm SU}(3)$, 
which connect nearest neighbor lattice sites. 
The gauge action (\ref{Sgauge}) is obtained by summing over the 
plaquette variables $U_{x,\mu \nu}$, given in eq.\ (\ref{plaq}).
Integrating out the quark fields $\bar \Psi$, $\Psi$
gives rise to the fermion determinant of eq.\ (\ref{fermdet}).
Thus the partition function takes the form
\be
Z = \int {\cal D}U \ {\rm det}\, M[U] \, \exp (-S_{\rm gauge}[U]) \ ,
\ee
{\it i.e.}\ the statistical weight is composed of the fermion
determinant and the Boltzmann factor of the gauge action, both depending
on the gauge configuration.

In lattice QCD, the prefactor of eq.\ (\ref{Sgauge}) is usually 
denoted as $\beta = 6/g^{2}$ (as in Statistical Mechanics).
If it increases, the ``golden'' gauge configurations 
(thermalized and decorrelated) become smoother, and the physical
lattice spacing $a$ decreases; $\beta \to \infty$ corresponds to
the continuum limit $a \to 0$. At fixed $\beta$ one selects a 
phenomenologically known, dimensional quantity, which is rather easy 
to measure numerically, and identifies $a$ in this way\cite{Rainer}, 
usually in the range $a \simeq 0.05 \dots 0.1~{\rm fm}$.
One would like $a$ to be small, 
to suppress the lattice artifacts, but on the other hand
the physical size $L a$ should be kept large enough 
(with the affordable number of lattice sites $L$ in one direction); 
it should clearly exceed the correlation length (cf.\ Section 2), 
which is given here by the inverse pion mass, $\xi = 1/ M_{\pi}$.

Once the Monte Carlo algorithm has generated a set of
``golden'' configurations with the right probability distribution, 
we can measure physical observables.
For instance, the pseudo-scalar densities (cf.\ footnote \ref{fermdense})
\be
P_{t}^{+} = \sum_{\vec x} \bar d_{\vec x,t} \, \gamma_{5} \, u_{\vec x,t} \ ,
\quad P_{t}^{-} = \sum_{\vec x} \bar u_{\vec x,t} \, \gamma_{5} \, d_{\vec x,t}
\ee 
(where $u$ and $d$ are the quark fields for the corresponding flavors) 
are relevant for the {\em charged pions}. $M_{{\pi}^{\pm}}$ is obtained 
from the decay of the correlator (at $1 \ll t/a \ll L_{t}$)
\be  \label{pidec}
C_{{\pi}^{\pm}} (t) = 
\la \, P^{\pm}_{s+t} \, P^{\pm \, *}_{s} \, \ra_{c} \propto 
\exp (- M_{\pi}^{\pm} \, t) \ ,
\ee
up to artifacts that we discussed in Section 2.
(Of course, CPT invariance implies $M_{\pi^{+}} = M_{\pi^{-}}\,$.)

Varying the density $P_{t}^{\pm}$, both with respect to the quark flavors
and the element of the Clifford algebra, leads to further hadron
masses (this is nicely described in the lecture notes \refcite{Marc}).
The same procedure allows us to measure further physical 
quantities, such as matrix elements, decay constants, the chiral
condensate, the crossover temperature of the chiral --- or 
confinement--deconfinement --- transition, the topological 
susceptibility etc. These results are {\em truly} based on first
principles of QCD (in contrast to many other approaches in the
literature, which are QCD-related, but which ultimately 
depend on additional assumptions and parameters).

Of course, lattice simulations can also be applied to other models
of (phenomenological or theoretical) interest, such as the Higgs sector 
of the Standard Model (based on triviality, an upper bound for 
the Higgs particle mass could be estimated, $M_{\rm Higgs} \ltapprox
650 \, {\rm GeV}$), QED (although in that case perturbation theory 
is successful), and many models which are relevant {\it e.g.}\ in 
Condensed Matter Physics, Statistical Mechanics, or the search for 
physics beyond the Standard Model\cite{prospects}, 
including multi-flavor systems (which are fashion now, in view
of a possible sub-structure of the Higgs particle\cite{Anna}), 
Dark Matter models\cite{Dark}, supersymmetry 
(although its lattice formulation is problematic\cite{Alessandra}), 
field theories in non-commutative spaces,\cite{4dNCU1} etc.

\begin{figure}[h!]
\vspace*{-4mm}
\includegraphics[width=6.2cm,angle=0]{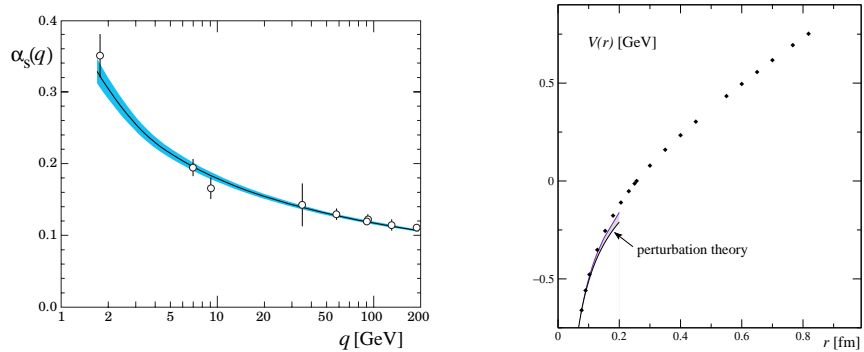}
\hspace*{1cm}
\includegraphics[width=4.4cm,angle=0]{pot.eps}
\caption{On the left: the strong coupling $\alpha_{s}$ of eq.\ (\ref{alphas})
as a function of a transfer momentum $q$. The curve logarithmically
interpolates phenomenological data points, and illustrates asymptotic 
freedom. On the right: the effective potential $V(r)$ for two static
quarks, separated by a distance $r$. At 
$r \protect\gtapprox 0.5\ {\rm fm}$, $V(r)$ increases linearly
(plots adopted from Ref.\ \protect\refcite{MLlect}.)}
\label{QCDplots}
\vspace*{-4mm}
\end{figure}

However, QCD simulations are particularly motivated, so let us
focus on QCD. Fig.\ \ref{QCDplots} (left) shows
phenomenological data which illustrate the famous property of
asymptotic freedom, {\it i.e.}\ the decrease of the strong
gauge coupling $g_{s}$, and the corresponding quantity $\alpha_{s}$,
for increasing transfer momentum $q$,
\be  \label{alphas}
\alpha_{s}(q) = \frac{1}{4 \pi} g_{s}(q)^{2} =
\frac{6 \pi}{33 - 2 N_{f}} \,
\frac{1}{\ln (q/\Lambda_{\rm QCD})} \quad \ {\rm (to~leading~order).}
\ee
We see that perturbation theory is applicable only at high energy,
where $\alpha_{s}(q)$ becomes small, but at low energy a non-perturbative
method is required.

The value of $\Lambda_{\rm QCD}$ is renormalization scheme dependent,
typically it is obtained in the range $200 \dots 250 \, {\rm MeV}$.
It is remarkable that an intrinsic scale occurs;
for massless quarks, the action does not
involve any dimensional parameter ($g_{s}$ is dimensionless).
Hence this scale is due to an anomaly:
quantization explicitly breaks the scale
invariance of the classical field theory. 
It is similar to the energy scale of the chiral condensate,
which was measured in lattice simulations,
\be  \label{chicon}
\Sigma = - \la \bar \Psi \, \Psi \ra \ , \quad
\Sigma^{1/3} \approx 270  \ {\rm MeV} \ .
\ee
$\Sigma$ acts as an order parameter for chiral symmetry breaking
(we refer to 2 or 3 light quark flavors at low temperature).
Thus QCD sets a magnitude for the light hadron masses.

The plot in Fig.\ \ref{QCDplots} (right)
illustrates confinement: 
$V(r)$ is the energy that it takes to pull apart two static
quarks by a distance $r$. At short distances it can be computed 
perturbatively (few gluon exchange), but the linear increase at 
$r\gtapprox 0.5\ {\rm fm}$ is measured on the lattice;
up to some $r$, it agrees with experimental observations.\footnote{In 
the real (dynamical) world, additional quark--anti-quark pairs 
are generated before $r$ becomes really large. They form new hadrons,
so confinement is actually more complicated. This (gluonic) ``string 
breaking'' has been studied on the lattice as well.\cite{stringbreak}}

\subsection{Hadron masses}

We pointed out in Subsection 2.5 that the fermion determinant is
particularly tedious to deal with; it is the bottleneck for the
generation of ``golden QCD configurations''. Therefore, in the 
last century it was ignored in many simulations; ${\rm det}\ M[U]$ was
treated as a constant, only $\, \exp (-S_{\rm gauge}[U]) \,$ was taken 
into account. This simplification is known as the {\em quenched
approximation,} and there are several ways to express its meaning:
in case of $N_{f}$ degenerate quark flavors,
${\rm det}\, M[U]$ contains the factor $({\rm det}\, D[U])^{N_{f}}$,
where $D[U]$ is the Dirac operator for one of these flavors.
In this sense, the quenched approximation corresponds to $N_{f}=0$.
Alternatively, we may refer to valence quarks with an infinite mass,
which also leads to ${\rm det}\, M[U] = {\rm const}$. This
shows that sea quarks are not included in this approximation.

\begin{figure}[h!]
\begin{center}
\vspace*{-2mm}
\includegraphics[width=6cm,angle=0]{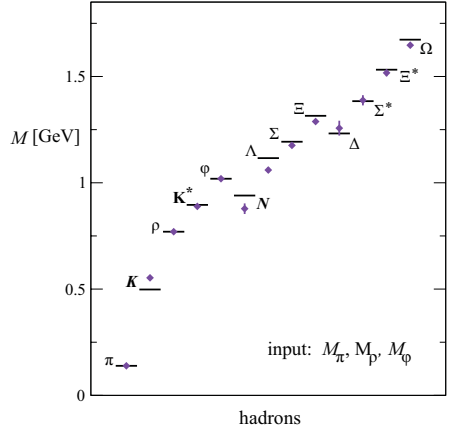}
\vspace*{-5mm}
\end{center}
\caption{The light hadron spectrum obtained by the CP-PACS 
Collaboration\cite{CPPACS} from lattice QCD in the quenched approximation.
The masses agree with the phenomenological values (horizontal bars) 
up to ${\cal O}(10) \, \%$
(plot adopted from Ref.\ \protect\refcite{MLlect}, in a modified form).}
\label{CPPACSfig}
\vspace*{-4mm}
\end{figure}

So quenched simulations were using an incomplete statistical weight 
for the configurations. This implies a systematic error, which 
is hard to estimate. One could worry that it distorts
the results completely, but the outcome was not that bad at all.
Fig.\ \ref{CPPACSfig} gives an overview of hadron masses, obtained 
quenched by the CP-PACS Collaboration\cite{CPPACS}.
Typically three input quantities are involved. We mentioned before
that measuring one phenomenologically known, dimensional quantity 
sets an overall scale.
The bare light quark masses are assumed to be degenerate,
$m_{u} = m_{d}$, and tuned for a suitable value of $M_{\pi}$, and some 
strange hadron mass fixes the bare quark mass $m_{s}$.
This set of hadrons does not include heavier valence quarks 
($c$, $b$, $t$), and their sea quark contribution is negligible at
low energy (up to a minor shift in the short-distance 
coupling).\footnote{On the other hand, $s, \, \bar s$ sea quarks 
contribute {\it e.g.}\ to the nucleon mass and spin at percent 
level.\cite{nucstrange}}
For the results of Fig.\ \ref{CPPACSfig}, these three input terms are
the masses $M_{\pi}$, $M_{\rho}$ and $M_{\phi}$, and all the rest
are lattice predictions. We see that the agreement with the 
phenomenological values (horizontal bars) is quite good,
it matches up to ${\cal O}(10) \, \%$. This is even more impressive
when we consider that --- for technical reasons --- quenched QCD
simulations were often performed at heavy pion masses, like
$M_{\pi} \gtapprox 600 \, {\rm MeV}$,\footnote{The $M_{\pi}$ values 
in Ref.\ \refcite{CPPACS} were not that heavy, in the range
$200 \dots 500  \, {\rm MeV}$.}
followed by a quasi-chiral extrapolation 
to the physical value $M_{\pi} \approx 138 \, {\rm MeV}$,
which caused additional uncertainties. \\

Hence quenched simulation results already indicated that QCD
is promising to work also at low energy.
Of course, the lattice community wanted to proceed to precise 
large-scale simulations with dynamical quarks. This was achieved 
in the 21st century, with rapid progress based on improvements 
in many respects (computers, algorithms, lattice actions,
numerical measurement techniques etc.). 

\begin{figure}[h!]
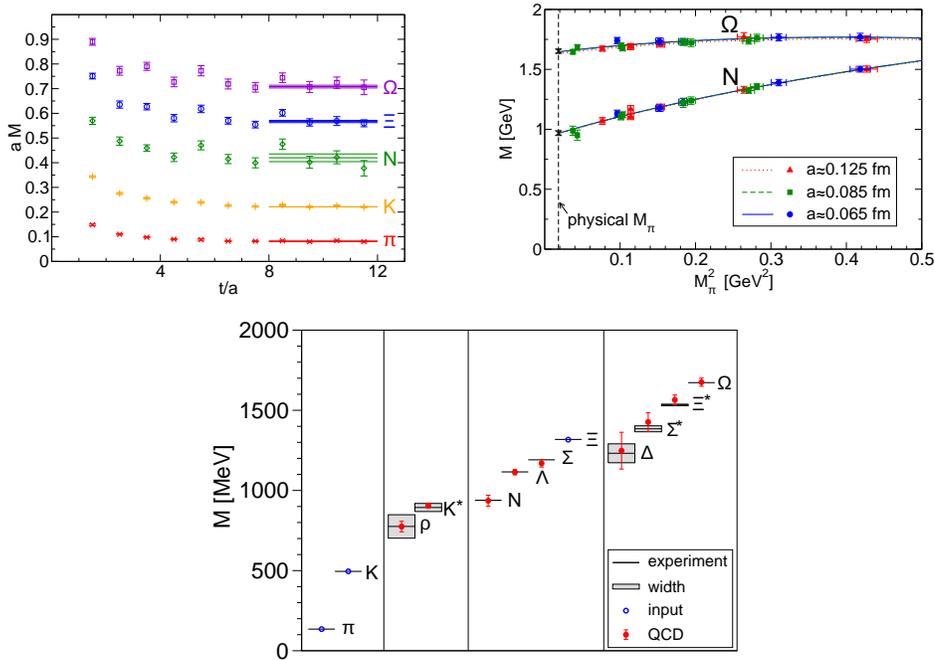

\vspace*{-2mm}
\begin{center}
\includegraphics[width=5.3cm,angle=0]{massplat.eps}
\hspace*{1cm}
\includegraphics[width=5.8cm,angle=0]{scaling.eps} \vspace*{3mm} \\
\includegraphics[width=7cm,angle=0]{spect.eps}
\end{center}
\vspace*{-1mm}
\caption{The hadron spectrum, obtained by the Budapest-Marseille-Wuppertal
Collaboration by QCD simulations with $2+1$ flavors of
dynamical quarks.\cite{BMW} 
On the left: illustration of the mass plateaux, which provide 
reliable values for a set of hadron masses, cf.\ eq.\ (\ref{Mht}). 
On the right: extrapolation of two baryon masses to the physical 
point, where $M_{\pi}$ takes its phenomenological value. Below: 
overview over the extrapolated hadron spectrum, compared to the 
experimental results (plots adopted from Ref.\ \protect\refcite{BMW}).}
\label{BMWplot}
\vspace*{-4mm}
\end{figure}

Fig.\ \ref{BMWplot} shows results by the Budapest-Marseille-Wuppertal
Collaboration\cite{BMW}. If we assume an exponential decay of a
hadronic correlation function $C$, as in the example of eq.\ 
(\ref{pidec}), we can extract the corresponding hadron mass $M$ as
\be  \label{Mht}
M = \frac{1}{a} \ln \Big( \frac{C_{t}}{C_{t+a}} \Big) \ .
\ee
In practice the results depend on the distance $t$.
The exponential that we refer to is actually just the leading 
contribution to a superposition of exponentials, which includes
excited states, so at short $t$ contaminations by higher states
are expected. As $t$ grows, they are exponentially suppressed
and the first energy gap dominates, which suggests to take
$t$ as large as possible. One should not overdo it, however,
because at large $t$ the wanted signal
is  exponentially suppressed too (though in a weaker form), 
and it eventually disappears in the statistical 
noise. Hence one hopes to find a suitable interval of moderate $t$,
where $M$ is stable and conclusive. Such intervals can
be observed in the first plot of Fig.\ \ref{BMWplot} (left), 
for five hadron masses.

A set of pion masses was simulated, which reached down to 
$M_{\pi} \approx 190 \, {\rm MeV}$, {\it i.e.} quite close to the
physical value.\footnote{Meanwhile simulations are performed
{\em at} the physical pion mass, 
($M_{\pi^{\pm}} \simeq 140 \, {\rm MeV}$,
$M_{\pi^{0}} \simeq 135 \, {\rm MeV}$),
or even below for a safe interpolation,
which eradicates the error source due to the ``chiral
extrapolation''.} Fig.\ \ref{BMWplot} (right) shows how two 
baryon masses depend on $M_{\pi}^{2}$, and how they are extrapolated
to the physical pion mass, which is also here an input parameter.

The system size was kept at $La \simeq 4/M_{\pi}$, so it
attained $\approx 4 \, {\rm fm}$, which suppresses
finite size effects quite well. Regarding the lattice artifacts,
three different lattice spacings were used as a basis for 
the continuum extrapolation, $a = 0.125\, {\rm fm},
\ 0.085\, {\rm fm}, \ 0.065\, {\rm fm}$.
In this way, the spectrum in Fig.\ \ref{BMWplot} (bottom) was
obtained, in excellent agreement with phenomenology.
Referring to Section 1, we are most interested in the nucleon
mass, which was measured as
\be
M_{\rm N} = 936(25)(22) \, {\rm MeV} \ ,
\ee
in accurate agreement with Nature. The parentheses give the 
statistical and systematic errors, respectively, 
which sum up to $3.6\, \%$. \\

\begin{figure}[h!]
\vspace*{-4mm}
\begin{center}
\vspace*{-2mm}
\includegraphics[width=5.3cm,angle=0]{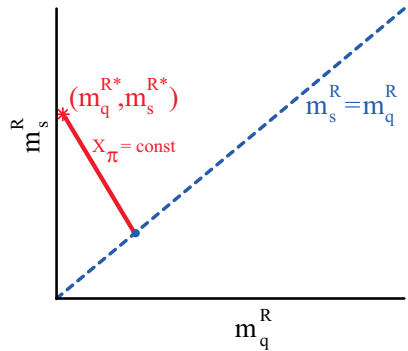}
\end{center}
\vspace*{-4mm}
\caption{Scheme of the renormalized quark masses 
$m_{q}^{R} := m_{u}^{R} =m_{d}^{R}$ and $m_{s}^{R}$.\protect\cite{QCDSFPLB} 
SU(3)$_{f}$ flavor symmetry is realized on the dashed line, where we 
select the point with the physical value of $X_{\pi}$. From there we decrease
$m_{q}^{R}$, and increase $m_{s}^{R}$, such that $X_{\pi}$ is preserved,
and we approach the physical masses $M_{\pi}$ and $M_{K}$.
This is the point to determine a multitude of further hadron masses.}
\vspace*{-4mm}
\label{quarkmasses}
\end{figure}

A new strategy was introduced in Ref.\ 
\refcite{QCDSFPLB}. The traditional approach for 2+1
flavors ($m_{u}=m_{d} \ll m_{s}$) proceeded as follows:
\begin{itemize}
\item Adjust the kaon mass, $M_{K} \simeq 496 \, {\rm MeV}$, as 
well as possible.
\item Now push for light pions, while keeping $M_{K}\approx {\rm const.}$ 
\end{itemize}
The new approach by the QCDSF-UKQCD Collaboration proceeds in a 
different manner, which is illustrated in Fig.\ \ref{quarkmasses}.
\begin{itemize}
\item Start from SU(3)$_{f}$ flavor symmetry, {\it i.e.}\
$m_{u} = m_{d} = m_{s}$ and $M_{\pi} = M_{K}$.
In this setting, the ``center of squared masses'' of the 
pseudoscalar meson octet,
\be
X_{\pi}^{2} := \frac{1}{3} (M_{\pi}^{2} + 2 M_{K}^{2}) \ ,
\ee
is tuned to its physical value.

\item Now decrease the light quark mass $m_{q} := m_{u}=m_{d}$
while increasing $m_{s}$.
This simultaneous modification is performed such that
$X_{\pi}$ is kept constant, while
$M_{\pi}$ and $M_{K}$ extrapolate to their physical values.

This trajectory towards the physical point is constrained and stable.
It is located inside the regime where Chiral Perturbation Theory
applies, which guides the extrapolation.
Flavor singlet quantities, like $X_{\pi}$, change only in the second
order of the variation of the renormalized quark masses.
\end{itemize}

\begin{figure}[h!]
\vspace*{-3mm}
\begin{center}
\includegraphics[width=9cm,angle=0]{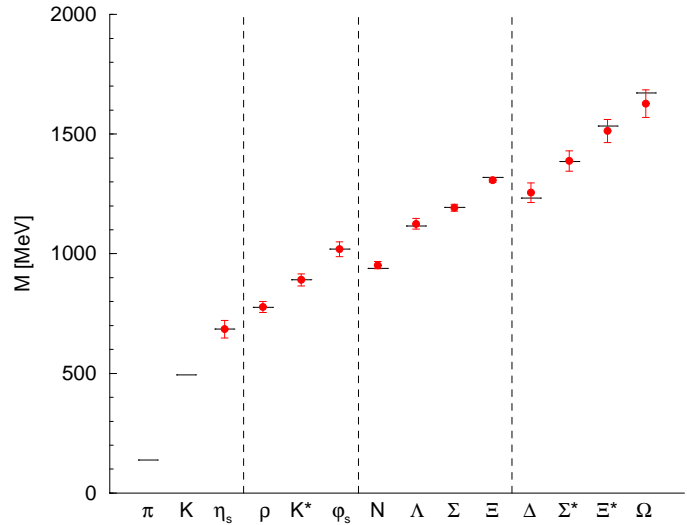}
\end{center}
\caption{The hadron spectrum obtained by the QCDSF-UKQCD
Collaboration\cite{QCDSFPRD} from QCD simulations with
$2+1$ flavors of dynamical quark. $M_{\pi}$ and $M_{K}$ are
input quantities, and all the rest agrees with phenomenology
within small errors (plot from Ref.\ \protect\refcite{QCDSFPRD},
in modified form).}
\label{QCDSFspectrum}
\vspace*{-4mm}
\end{figure}

The application of this method\cite{QCDSFPRD} led to the hadron 
spectrum in Fig.\ \ref{QCDSFspectrum}. It was obtained in the
lattice volumes $V= 24^{3} \times 48$ and $32^{3} \times 64$,
at a lattice spacing $a = 0.0765(15)~{\rm fm}$. We see an
accurate agreement with the phenomenological masses,
up to the baryon $\Omega \sim (sss)$.

Further collaborations have arrived at similar results.
In light of these data, no doubt remains that QCD is the 
appropriate theory for hadron physics down to low energy.
Overviews of the lattice results on light hadrons are given 
in Refs.\ \refcite{FodHoel} and \refcite{FLAG}; the latter contains
carefully evaluated {\em ``world averages''} for hadron masses,
and numerous additional quantities like quark masses, 
Low Energy Constants, decay constants, meson mixing parameters, 
form factors and the running coupling $\alpha_{s}$.
We also refer to the review talks at the annual Lattice Symposium;
since 2005 the proceedings appear in {\em Proc.\ of Science (PoS)}.

\subsection{Lattice QCD and chiral perturbation theory}

Before the breakthrough of lattice QCD, low energy QCD was
most successfully described by the effective Lagrangian of
Chiral Perturbation Theory. Unlike {\em ad hoc} effective models,
it is manifestly derived from QCD with light quark flavors.
In the limit of massless quarks, the left- and right-handed
quark field decouple,
\be
{\cal L}^{\rm QCD}_{m_{\rm quark}=0} = \bar \Psi_{L} D \Psi_{L} +
\bar \Psi_{R} D \Psi_{R} + {\cal L}_{\rm gauge} \ ,
\ee
where $\bar \Psi$, $\Psi$ run over $N_{f}$ flavors, and $D$ 
is the (joint) Dirac operator. Thus ${\cal L}^{\rm QCD}_{m_{\rm quark}=0}$ 
has a global $U(N_{f})_{L}\otimes U(N_{f})_{R}$ symmetry, 
which can be decomposed into subgroups,
\be
{\rm U}(N_{f})_{L} \otimes {\rm U}(N_{f})_{R} =
{\underbrace{{\rm SU}(N_{f})_{L} \otimes {\rm SU}(N_{f})_{R}}
_{\rm chiral~flavor~symmetry}} \otimes
{\underbrace{{\rm U}(1)_{L=R}}_{\rm baryon~number}}
\otimes{\underbrace{{\rm U}(1)_{\rm axial}}_{\rm anomalous}} \ .
\ee
One U(1) symmetry breaks explicitly under quantization (axial anomaly). 
The other one is chirality-blind and responsible for 
baryon number conversation. We are interested in the remaining 
{\em chiral flavor symmetry}, which breaks spontaneously,
\be
{\rm SU}(N_{f})_{L} \otimes {\rm SU}(N_{f})_{R} \to {\rm SU}(N_{f})_{L=R} \ .
\ee
This yields $N_{f}^{2}-1$ Nambu-Goldstone Bosons (NGBs), or
light quasi-NGBs if we include small quark masses.
They dominate the low energy physics, and Chiral Perturbation Theory
deals with a Lagrangian for these quasi-NGBs, which are identified
with the light mesons (pions for $N_{f}=2$; $\pi,\, K,\, \eta$
for $N_{f}=3$).

Hence Chiral Perturbation Theory deals with 
fields in the {\em coset space of the spontaneous symmetry breaking,}
$U(x) \in {\rm SU}(N_{f})$.\cite{XPT} It includes all terms, which 
obey the chiral flavor symmetry (as well as Lorentz invariance
and locality).
They are structured in a hierarchical order, according to the energy
(powers of momenta, and of quark or meson masses).

Here we consider $N_{f}=2$, such that the field $U(x) \in {\rm SU}(2)$
describes the pion triplet $\pi^{+},\, \pi^{0},\, \pi^{-}$, and
we assume a (degenerate) mass $m_{q}$ for $u$ and $d$ quarks.
The leading and some sub-leading terms of the effective Lagrangian
can be written as
\bea
&& \hspace*{-8mm} 
{\cal L}_{\rm eff} (U, \partial_{\mu} U) =
\frac{F_{\pi}^{2}}{4} \, {\rm Tr}
[ \partial_{\mu} U^{\dagger} \partial_{\mu}U ]
+ \frac{\Sigma \, m_{q}}{2} \, {\rm Tr} [U^{\dagger} + U ] \nn \\
&& \hspace*{-8mm}  
+ \, l_{1} \ ({\rm Tr} [ \partial_{\mu} U^{\dagger} \partial_{\mu} U ])^{2}
+ l_{2} \ ({\rm Tr} [ \partial_{\mu} U^{\dagger} \partial_{\nu} U ] )^{2}
+ l_{3} \ \Big( \frac{\Sigma m_{q}}{F_{\pi}^{2}} \Big)^{2}
({\rm Tr} [ U^{\dagger} + U ])^{2} + \dots
\eea
The coefficients to these terms are known as {\em Low Energy
Constants} (LECs); they appear in the effective theory as free
parameters. Hence ${\cal L}_{\rm eff}$ {\em is} derived from QCD,
but --- being a simplification --- it cannot capture its 
entire information.

The leading order LECs are the {\em pion decay constant}
$F_{\pi}$ and the {\em chiral condensate} $\Sigma$, while the
$l_{i}$ appear in sub-leading terms. The LECs of the effective
theory can only be determined from the underlying, fundamental theory,
which is QCD in this case. Since this refers to low energy, it is
a non-perturbative problem, and therefore a challenge for lattice
simulations. If this works, we arrive at a rather comprehensive
picture of low energy QCD.

Chiral Perturbation Theory has been formulated and studied 
in various regimes, depending on the volume:
\begin{itemize}
\item $p$-regime\cite{preg}: the volume is large, 
{\it e.g.}\ $V = L^{4}$, $L \gg M_{\pi}^{-1}$,
and the momentum of $U(x)$ counts (in the energy hierarchy) 
like $M_{\pi}$.
\item $\epsilon$-regime\cite{epsreg}: the volume is small,
$L \ltapprox M_{\pi}^{-1}$, and a momentum counts like $m_{q}$.
\item $\delta$-regime\cite{deltareg}: the {\em spatial} volume 
is small, but the extent in Euclidean time is large, {\it e.g.}\
$V = L^{3} \times L_{t}$, $L \ltapprox M_{\pi}^{-1} \ll L_{t}$.
\end{itemize}

The $p$-regime is standard, and also the $\epsilon$-regime has
been studied extensively, both analytically\cite{LeuSmi} and 
numerically. The LECs are the same in all three regimes, hence the 
$\epsilon$-regime is convenient for their determination
by simulations, which do not require large volumes. Since
chirality is vital in this context, it is appropriate to
use Ginsparg-Wilson fermions for the quarks. This
is computationally expensive, so it was done in the quenched
approximation even in this century, which led to decent 
results for the leading LECs. For instance, Random Matrix Theory 
predictions\cite{Poul} for the low lying Dirac eigenvalues match the
numerical data well, and the fit determines $\Sigma$.\cite{RMT,Stani}
$F_{\pi}$ is easier to extract from the correlation of axial 
currents,\cite{AA,Stani} or of pseudo-scalar zero modes.\cite{0modes,Stani}
Recent progress, now with dynamical quark simulations, 
is reviewed in Ref.\ \refcite{Giusti}.

\begin{figure}[h!]
\vspace*{-2mm}
\begin{center}
\includegraphics[width=11cm,angle=0]{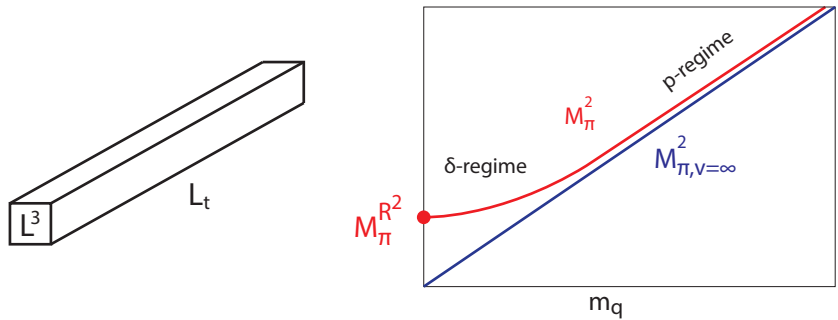}
\end{center}
\vspace*{-4mm}
\caption{A schematic illustration of a volume, which corresponds 
to the $\delta$-regime (left) and of the quark mass dependence
of $M_{\pi}^{2}$: at large $m_{q}$ it approaches $M_{\pi}^{2} \propto m_{q}$, 
whereas the chiral limit leads to the residual pion mass $M_{\pi}^{R}$.}
\vspace*{-4mm}
\label{deltavol}
\end{figure}

Here we discuss the least explored case, the {\em $\delta$-regime}
(for a recent study, see Ref.\ \refcite{Tibur}).
The corresponding volume is illustrated in Fig.\ \ref{deltavol}
(left). Its quasi-1d form enables an approximate analytical 
treatment as a quantum mechanical system, {\it i.e.}\ a 1d
O(4) model\cite{deltareg} (due to the local isomorphy 
${\rm O}(4) \sim {\rm SU}(2)\otimes {\rm SU}(2)$).
Since the spatial volume is finite, there is no spontaneous 
symmetry breaking, and pions do {\em not} become massless NGBs
at $m_{q} \to 0$. Instead they keep a {\em residual mass} $M_{\pi}^{R}$
in the chiral limit. On the other hand, at large $m_{q}$ the 
same volume appears large, and we obtain the $p$-regime behavior, 
namely the Gell-Mann--Oakes--Renner Relation 
$M_{\pi}^{2} \approx (\Sigma /F_{\pi}^{2}) \, m_{q}$.
This is shown schematically in Fig.\ \ref{deltavol} (right).

We return to the picture of an O(4) rotor: the first gap in
its energy spectrum
$E_{n} = n(n+2)/(2 \Theta)$ ($n=0,1,2\dots$) determines
the residual pion mass, $M_{\pi}^{R}= 3/(2 \Theta)$.
The question is how to identify $\Theta$, the moment of inertia.
An expansion in inverse powers of $(F_{\pi}L)^{2}$ yields
\be
L M_{\pi}^{R} = \frac{3}{2 (F_{\pi} L)^{2} (1 + \Delta )} \ , \quad
\Delta = \frac{0.452\dots}{(F_{\pi} L)^{2}} + 
{\cal O} \Big( \frac{1}{(F_{\pi} L)^{4}} \Big) \ .
\ee
The first order ($\Delta =0$) was derived in Ref.\ \refcite{deltareg}, 
and the second order in Ref.\ \refcite{HasNie}. For the third order 
Refs.\ \refcite{Has10,NieWei} obtained slightly different results,
but for sure at this level some sub-leading LECs $l_{i}$ enter.

\begin{figure}
\begin{center}
\includegraphics[angle=270,width=.6\linewidth]{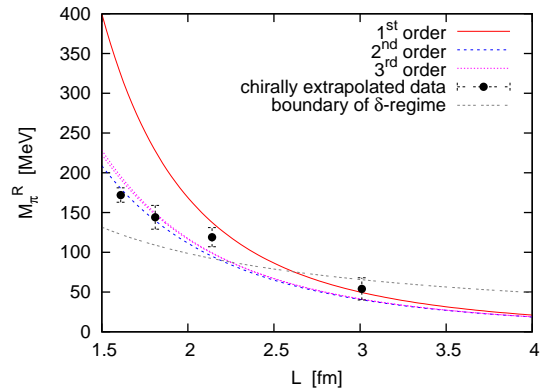}
\end{center}
\vspace*{-2mm}
\caption{The residual pion mass in the $\delta$-regime of QCD.
We show the predictions of the first\cite{deltareg}, 
second\cite{HasNie} and third\cite{Has10} order $\delta$-expansion; 
the latter two are close, which indicates convergence.
The data points are extrapolated numerical results by the 
QCDSF Collaboration\cite{QCDSFdelta}, which agree well
with the quasi-converged $\delta$-expansion (plot from the
second article cited in Ref.\ \protect\refcite{QCDSFdelta}).}
\label{MpiR}
\vspace*{-3mm}
\end{figure}

The QCDSF Collaboration performed extrapolations of 
numerical pion mass measurements into the $\delta$-regime\cite{QCDSFdelta}, 
and obtained good agreement with these predictions, see Fig.\ \ref{MpiR}.
In the sub-leading order of the effective Lagrangian, 
the least known --- and most controversial LEC --- is $l_{3}$.
Its value is usually taken at the energy of the physical
pion mass, where it is denoted as $\bar l_{3}$. If we fix the other 
LECs involved to their known values, the QCDSF numerical results
suggest $\bar l_{3} = 4.2(2)$, somewhat above the world-average 
according to the FLAG report\cite{FLAG}, \ $\bar l_{3}|_{N_{f}=2} = 3.41(41)$.

\section{Status and challenges of the future}

Regarding the light hadron spectrum, low energy QCD is now
systematically tested from first principles: lattice simulations 
consistently confirm the phenomenological values up to ${\cal O}(1)\, \%$ 
uncertainty. This was a key point in the ambitious program, 
which was outlined in the 1980s. 
In particular, the precise results for the nucleon mass explain 
over $99 \, \%$ of the mass of known matter in the visible Universe.

Of course, lattice QCD results involve {\em many} quantities, 
beyond hadron masses and Low Energy Constants,
which were not discussed here:
decay constants, matrix elements, quark masses and the strong
coupling $\alpha_{s}$, flavor mixing parameters, thermodynamic 
properties, ingredients of structure functions, spin and (anomalous) 
magnetic momenta, quark and gluon propagators, vertex functions, etc.
Here we sketched the basic ideas --- they are explained in
detail in six text books\cite{books} devoted to this subject ---
and we showed a few selected examples; for a comprehensive overview 
of results (with light quarks) we refer again to Ref.\ \refcite{FLAG}.

We mentioned before that --- in the early period --- Kenneth Wilson 
was the main protagonist of the lattice approach to QCD. 
At the end of the 1980s, however, he suddenly expressed his pessimism 
about its prospects in any foreseeable future, and he left the field.
However, the lattice community kept working and growing:
the annual Lattice Symposium started as a small event in the
1980s, and attracted already some 300 participants in the 1990s.
Nowadays the number is around 500, and the total community
is certainly more than twice that large. Unfortunately, Latin
America is still drastically under-represented, we hope for that
to change in the future. Actually it is becoming easier to 
contribute even without huge computational resources, thanks to 
the {\em International Lattice Data Grid,}\cite{ILDG} which
makes valuable QCD configurations publically available.

Regarding Wilson's pessimism, we add that indeed it took 
time to arrive at percent level results for phenomenological 
quantities, but this has now been achieved.

So what is next? Meanwhile the lattice community is already pushing
for the sub-percent level. This is not a straightforward extension,
but it requires qualitatively new aspects: for instance QED effects 
and the splitting between the masses 
$m_{u}$ and $m_{d}$ have to be considered at that level of precision.
As a remarkable result, even the splitting between the neutron and
proton mass has been demonstrated\cite{Borsanyi15}:
it was shown that QCD implies in fact $M_{\rm neutron} > M_{\rm proton}$
--- a property, which is highly non-trivial, with far-reaching 
consequences like the {\em proton stability.}
Moreover the scenario with a quark mass $m_{u}=0$ --- which would
have solved the {\em strong CP problem}\footnote{It can be considered
natural to add a term $-\ri \theta Q[U]$ to $S_{\rm QCD}$, where $Q \in \Z$
is the topological charge (cf.\ Appendix A), 
and $\theta \in (-\pi, \pi]$ is the {\em vacuum angle.}
However, from the electric dipole moment of the neutron we infer that
this term seems to vanish ($|\theta | \ltapprox 10^{-10}$)\cite{PDG}. 
The {\em strong CP problem} is the quest for an explanation for this
peculiar value, which preserves CP invariance. 
If the mass of (at least) one quark flavor vanishes, then the
gauge configurations with $Q[U] \neq 0$ do not contribute to the
functional integral: according to the Atiyah-Singer Index Theorem,
the Dirac operator has zero modes for these configurations, 
which imply ${\rm det} \, M =0$, so the puzzle would 
have been solved.\label{CPprob}} --- could be safely ruled 
out.\cite{FLAG,munonzero}

Nevertheless there are still many outstanding challenges.\cite{prospects}
For instance, the masses of {\em excited states} are far more difficult 
to extract, hence their uncertainties are much larger. A notorious
example is the Roper resonance, $N^{*}(1440)$, where most lattice 
studies arrived at energies well above the phenomenological value
(compared to the negative parity state $N^{*}(1535)$, the masses
are often obtained in inverse order). The current status and recent 
progress are discussed in Ref.\ \refcite{Leinw}.

A general goal for the future is the step from post-dictions
(reproduction of experimentally known facts from QCD) to 
{\em predictions.} Indeed, an example was given already,
in the framework of {\em heavy quarks.} They are difficult to
handle on the lattice, due to the short Compton wavelength.
Nevertheless, in 2005 Ref.\ \refcite{HPQCD}
predicted the mass of the $B_{c}^{\pm}$ 
meson as $6.30(2)\, {\rm GeV}$, and one year later it was indeed 
measured experimentally\cite{CDF} at $6.286(5)\, {\rm GeV}$
(meanwhile the value slipped slightly down 
to $6.275(1)\, {\rm GeV}$).\cite{PDG}

So is everything going to continue smoothly, with 
straightforward progress and constant success?
This would by untypical for the history of science, and in fact
there are still severe conceptual obstacles ahead of us, which cannot
be overcome just by computer power.
In the appendices we comment on two examples.

\section*{Acknowledgements}

It is a pleasure to thank all my collaborators in the subjects
which have been addressed here, namely
I.\ Bautista,
V.\ Bornyakov, 
T.\ Chiarappa,
N.\ Cundy,
C.\ Czaban,
M.\ Dalmonte,
P.\ de Forcrand,
A.\ Dromard,
W.\ Evans,
U.\ Gerber,
M.\ G\"{o}ckeler,
L.\ Gonglach,
I.\ Hip,
C.P.\ Hofmann,
R.\ Horsley,
K. Jansen,
A.D.\ Kennedy,
C.\ Laflamme,
W.G.\ Lockhart,
H.\ Mej\'{\i}a-D\'{\i}az,
K.-I.\ Nagai,
Y.\ Nakamura, 
J.\ Nishimura,
H.\ Perlt,
D.\ Pleiter, 
P.E.L.\ Rakow, 
A.\ Sch\"{a}fer,
G.\ Schierholz,
A.\ Schiller,
S.\ Shcheredin,
T.\ Streuer,
H.\ St\"{u}ben,
Y.\ Susaki,
J.\ Volkholz,
M.\ Wagner,
U.-J.\ Wiese,
F.\ Winter,
J.M.\ Zanotti
and P.\ Zoller.
I am indebted to P.O.\ Hess for inviting me to contribute to this 
Focus Issue of the {\em Int.\ J.\ Mod.\ Phys.\ E}, and to 
A.\ Guevara and A.\ Rodr\'{\i}quez for their help with the figures.
This work was supported by the {\it Consejo Nacional de 
Ciencia y Tecnolog\'{\i}a} (CONACYT) through project 
CB-2010/155905, and by {\it DGAPA-UNAM,} grant IN107915.

\appendix

%\vspace*{-4mm}

\section{Numerical measurements at fixed topology}

In some field theory models, including QCD, the configurations occur 
in distinct topological sectors, labelled by a topological charge 
$Q\in \Z$. In the continuum, configurations can only be deformed 
continuously within the same sector (with periodic boundary
conditions, at finite action). 

Strictly speaking, there are no topological sectors on the
lattice (in most formulations). Nevertheless there are 
ways to define a topological charge $Q$ for the lattice
configurations. Continuous transitions are possible,
but they have to pass through a region of high Euclidean
action, which is statistically suppressed. As we make
the lattice spacing finer and finer, we approach more and more
the continuum behavior of infinite potential barriers. 
This means that an algorithm, which performs small update steps, 
tends to be blocked in one sector for a very long (computation) 
time, {\it i.e.}\ the auto-correlation time with respect to $Q$ 
gets {\em very} long.\cite{Rainer2}

This could in principle be overcome by an algorithm,
which performs large leaps to generate its Markov chain, 
like a cluster algorithm, but for lattice QCD no efficient algorithm
of this kind is known. So far, there is no actual competitor of the 
HMC algorithm, which proceeds in small update steps 
(regarding the Molecular Dynamics evolution).
Hence the problem of ``topological freezing'' is expected to 
become severe when we proceed to finer lattices.
Up to now, most simulations were performed at 
$0.05 \, {\rm fm} \ltapprox a \ltapprox 0.1 \, {\rm fm}$.
When we try to further suppress this source of systematic errors 
and work at $a \ltapprox 0.05 \, {\rm fm}$, the energy barriers
between the topological sectors increase and tunnelling between
them will be extremely rare.

\begin{figure}
\begin{center}
\includegraphics[angle=270,width=.47\linewidth]{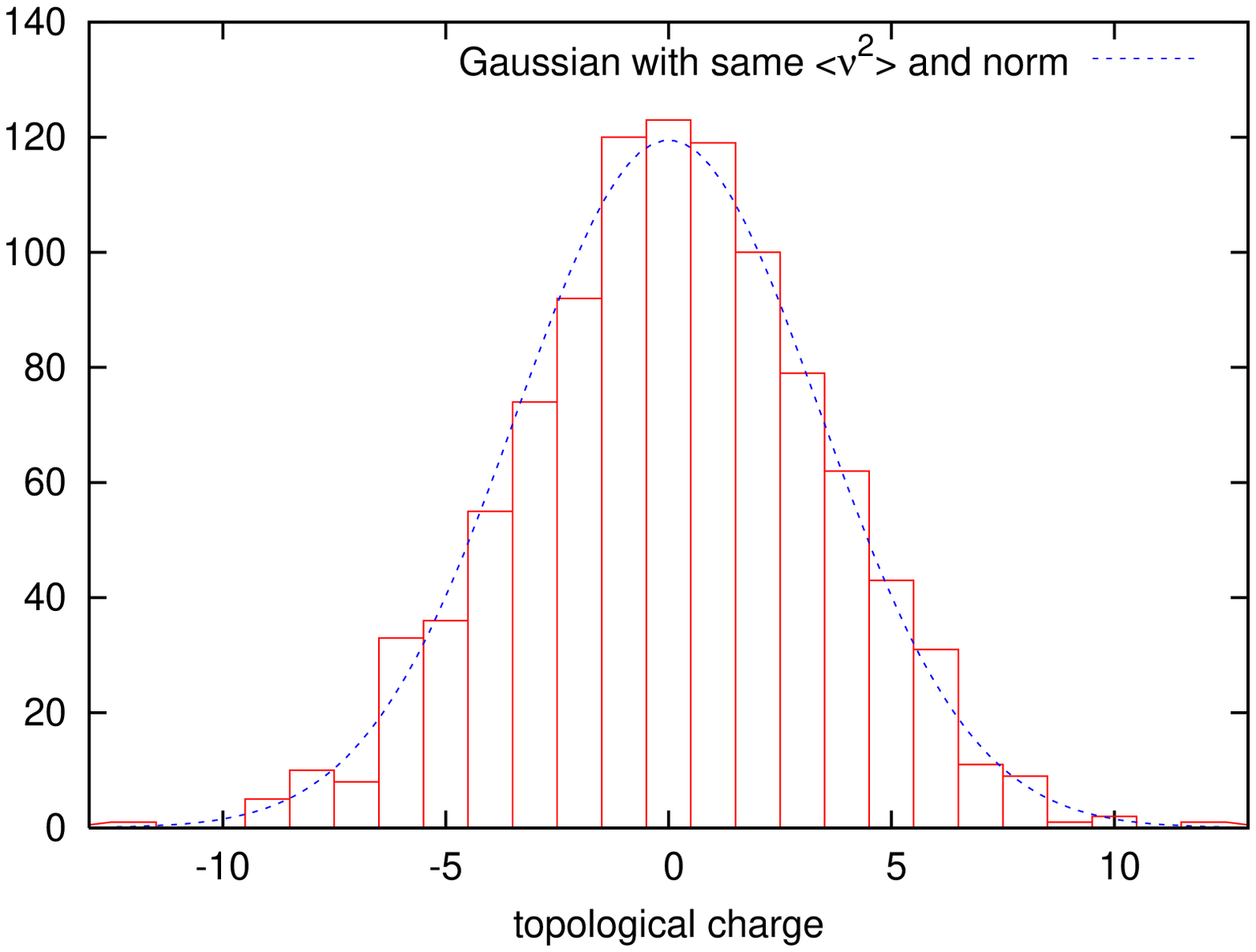}
\vspace*{3mm}
\includegraphics[angle=270,width=.47\linewidth]{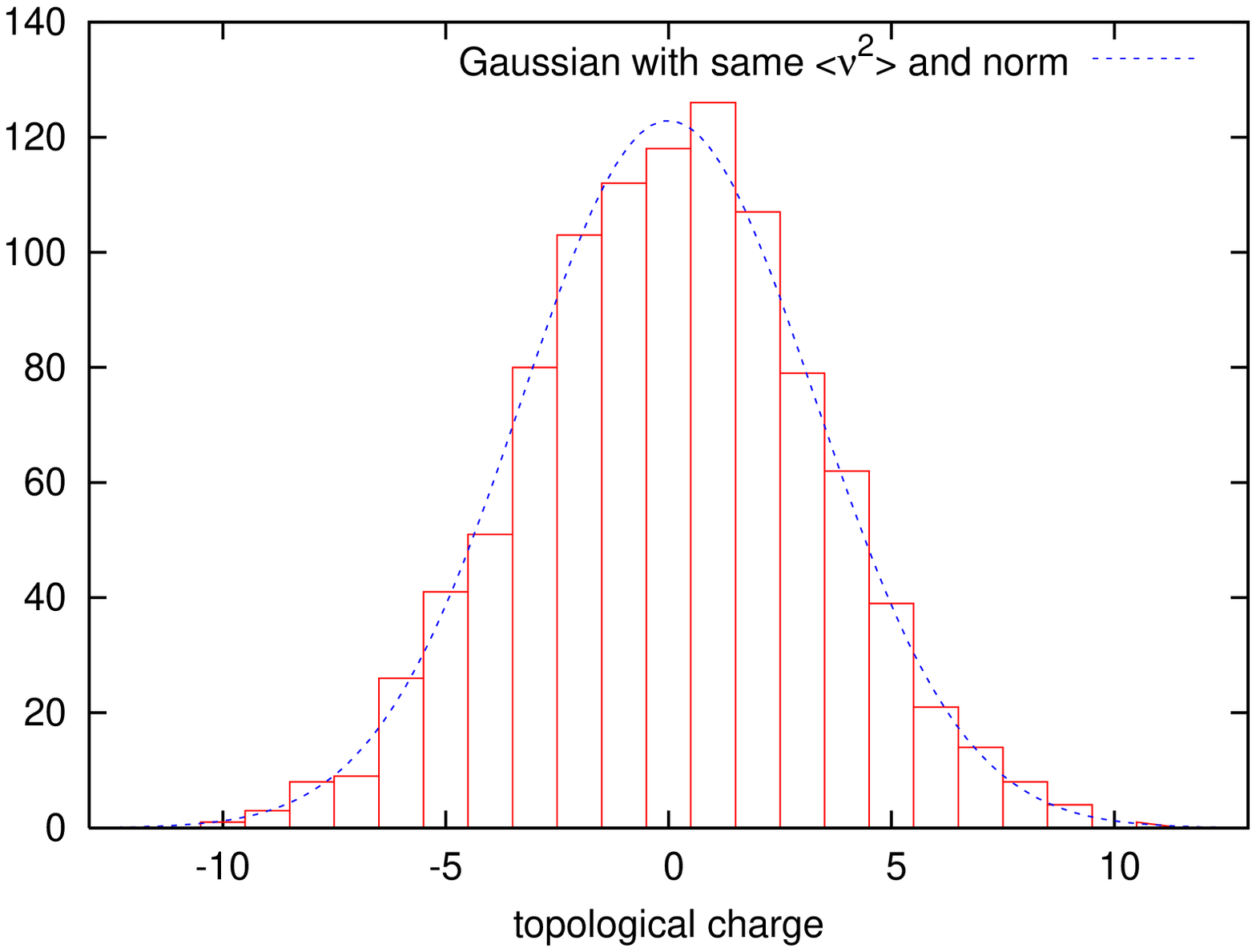}
\end{center}
\vspace*{-3.5mm}
\caption{Histograms for topological charges of lattice QCD
configurations\cite{Stani}, defined by the index of two variants of a
chiral lattice Dirac operator, (standard) overlap\cite{overlap} and 
(improved) hypercube-overlap\cite{ovHF}. 
These histograms are compatible with a
Gauss distribution, and its width determines the topological
susceptibility $\chi_{\rm t}$ 
(plots adopted from Ref.\ \protect\refcite{Stani}).}
\label{histo}
\vspace*{-2mm}
\end{figure}

This also depends on the (lightest) quark masses involved,
cf.\ footnote \ref{CPprob}.
For instance, in the quenched approximation, at moderate
lattice spacing, tunnelling happens quite easily, 
and one can measure {\it e.g.}\ the topological susceptibility
\be
\chi_{\rm t} = \frac{1}{V} \Big( \la Q^{2} \ra - \la Q \ra ^{2}\Big)
\ee
directly, or by a Gaussian interpolation of a topological charge histogram.
(In QCD parity symmetry holds, hence the second
term vanishes, $\la Q \ra = 0$.)
Examples for quenched QCD results,\cite{Stani} with $Q$ defined by the
index of the standard overlap Dirac operator\cite{overlap}, or
an improved version, the overlap-hypercube Dirac operator\cite{ovHF},
are shown in Fig.\ \ref{histo}. 
They are compatible with a Gaussian distribution. Its
width corresponds to a value of $\chi_{\rm t}$, which (roughly) supports
the {\em Witten-Veneziano formula}\cite{WiVe}. This formula relates the
mass of the $\eta'$-meson to the quenched value of $\chi_{\rm t}$, 
based on a $1/N_{c}$ expansion. Its simplest form reads
\be
M_{\eta'}^{2} \approx \frac{2 N_{f}}{F_{\pi}^{2}} \chi_{\rm t} \qquad
(M_{\eta'} \simeq 957.8\, {\rm MeV} , \ F_{\pi} \simeq 92.2\, {\rm MeV}) \ .
\ee
This formula explains the amazingly heavy $M_{\eta'}$
(``U(1) problem'') quantitatively, as a topological effect.\\

In view of this straight way to determine $\chi_{\rm t}$,
it seems questionable if this quantity can be measured based on 
data of a Monte Carlo history, which never --- or hardly ever ---
changes the topological sector. However, several methods for this
purpose were proposed, and recently also tested, and it turns out
to be possible under suitable conditions. In particular, the
Aoki-Fukaya-Hashimoto-Onogi formula\cite{AFHO}
\be  \label{AFHOeq}
^{\lim}_{|x|\to \infty} \ \la q_{0} \, q_{x} \ra_{|Q|} \simeq
- \frac{\chi_{\rm t}}{V} + \frac{Q^{2}}{V^{2}} + \dots
\ee
allows for an (approximate) determination of $\chi_{\rm t}$ based on the
correlation of the topological charge density $q_{x}$, measured
at fixed $|Q|$. Successful tests of this approach, 
and variants, are reported in Refs.\ \refcite{AFHOresults,Arthur16}.

An alternative approach divides the volume into sub-volumes 
(``slabs'') and determines $\chi_{\rm t}$  from the
topological charge distributions in these slabs. There the density
is summed up, $q = \sum_{x \in {\rm slab}} q_{x}$
($q$ does not need to be integer, since not all slab boundaries 
are periodic). Assuming the probability distribution
$p (q)$ to be Gaussian, we obtain for two slabs --- 
with volumes $xV$ and $(1-x)V$ --- in the sector with total 
topological charge $Q$,
\bea
p_{1} (q) \cdot p_{2} (Q-q)|_{Q} & \propto & \exp \Big(
- \frac{q^{2}}{2 \chi_{\rm t} x V }\Big) \
\exp \Big( - \frac{(Q-q)^{2}}{2 \chi_{\rm t} (1-x) V}\Big) \nn \\
& \propto & \exp \Big( - \frac{1}{2 \chi_{\rm t} V} 
\frac{(q - x Q)^{2}}{x (1-x)} \Big) \ .
\eea
Numerical data for $\la (q - x Q)^{2} \ra$, as a function of $x \in (0,1)$,
in a fixed sector $Q$, enable therefore the determination of 
$\chi_{\rm t}$, which was confirmed in tests with 
non-linear $\sigma$-models\cite{slab} and 2-flavor QCD\cite{Arthur16}.
For a related approach, see Ref.\ \refcite{LSD}.

Another method, which is specific to the simulation with
dynamical overlap quarks, was suggested in Ref.\ \refcite{Egri}.
This is the setting where the most extreme topological freezing has 
been observed\cite{topfreez}.\\

More generally, the question is how to measure
{\em any observable,} if the Monte Carlo histories are 
trapped in one topological sector.
Ref.\ \refcite{Liao} suggests to gradually fill the potential
valleys inside the topological sectors, such that transitions occur. 
Thus one simulates with modified Boltzmann probabilities, to be
corrected at the end by reweighting.

Alternatively, it has been advocated to prevent ``topological freezing'' 
by the use of partially {\em open boundary conditions}\cite{openbc}.
Since this breaks lattice translation invariance, also a milder form 
was proposed (a parity flip at the boundary)\cite{Mages}. 
In these scenarios, $Q$ is not integer anymore, so it
can change continuously.

However, it would be nicer to maintain periodic boundaries, and 
therefore $Q \in \Z$. Indeed, there is hope for an (approximate) 
determination of an expectation value of some observable, 
$\la \Omega \ra$, even if only topologically restricted measurements,
{\it i.e.}\ results for $\la \Omega \ra_{|Q|}$, 
are available. Brower, Chandrasekharan,
Negele and Wiese derived the approximation\cite{BCNW}
\be  \label{BCNWeq}
\la \Omega \ra_{|Q|} \approx \la \Omega \ra + \frac{c}{V \chi_{\rm t}}
\Big( 1 - \frac{Q^{2}}{V \chi_{\rm t}} \Big) \ , \quad (c = {\rm const.}) \ .
\ee
In an extremely large volume $V$, any sector provides the
same --- correct -- result. However, in moderate volumes, where
simulations are more realistic, a set of results for the 
left-hand-side, in various $V$ and $|Q|$, and a 3-parameter fit 
allow for the determination of the unknown terms 
$\la \Omega \ra$, $\chi_{\rm t}$ and $c$, where the former two are
of physical interest. These three terms are $V$-independent,
up to ordinary finite size effects, which tend to be exponentially
suppressed (cf.\ Section 2). In contrast, the topologically 
restricted quantities suffer from polynomial finite size effects,
as eqs.\ (\ref{AFHOeq}) and (\ref{BCNWeq}) show.

The latter is the beginning of an expansion in $1/(V \chi_{\rm t})$, 
which has recently been extended,\cite{Frankfurt} 
and even ordinary finite size effects have been included\cite{actpol}.
Detailed tests\cite{BCNWtest} in four models confirm that this 
method can provide accurate results for $\la \Omega \ra$ under
suitable conditions: $V \chi_{\rm t} = \la Q^{2}\ra > 1$ and
sectors with small $|Q|$ ($|Q| \leq 1$, or perhaps $|Q| \leq 2$). 
For the determination of $\chi_{\rm t}$,
however, the approaches of Refs.\ \refcite{AFHO,AFHOresults,slab}
provide better results.

\section{Quantum simulations as a remedy to the sign problem?}

Let us finally address the {\em phase diagram of QCD.}
Finite temperature QCD is difficult to explore,
since the Euclidean time direction has to be much shorter
than the spatial direction, say $V = L^{3} \times L_{t}$,
$L_{t} \ll L$. Let's consider lattice units: if 
the computational resources set a limit like $L=32$, 
for example, then $L_{t}$ must be really short (in the past
it was often just $L_{t} =4$). Converting it into a temperature 
of interest (now in physical units, with $k_{B}=1$),
$T = 1/(a L_{t})$, requires a large physical lattice spacing
$a$, which leads to bad lattice artifacts.

Hence the finite temperature behavior has been a challenge
for a long time, and the results were controversial. However,
over the last decade also this issue has been settled quite well,
thanks to improved simulations.\cite{Ludmilla} A confinement--deconfinement
crossover was observed, which seems to coincide with the chiral 
symmetry breaking/restoration transition (we mentioned the chiral 
condensate $\Sigma$ of eq.\ (\ref{chicon}) as the order parameter). 
It would be a phase transition for massless quarks, but since it is 
only a crossover, the exact transition temperature is somewhat 
criterion dependent; 
at $N_{f}=3$, values were obtained in the range\cite{Tcross}
\be
T_{\rm crossover} \simeq 150 \dots 160 \, {\rm MeV} \ .
\ee
(In this case, the quenched approximation deviates strongly,
$T_{\rm crossover}^{\rm quenched} \simeq 270 \, {\rm MeV}$.)\\

\begin{figure}
\includegraphics[angle=0,width=.41\linewidth]{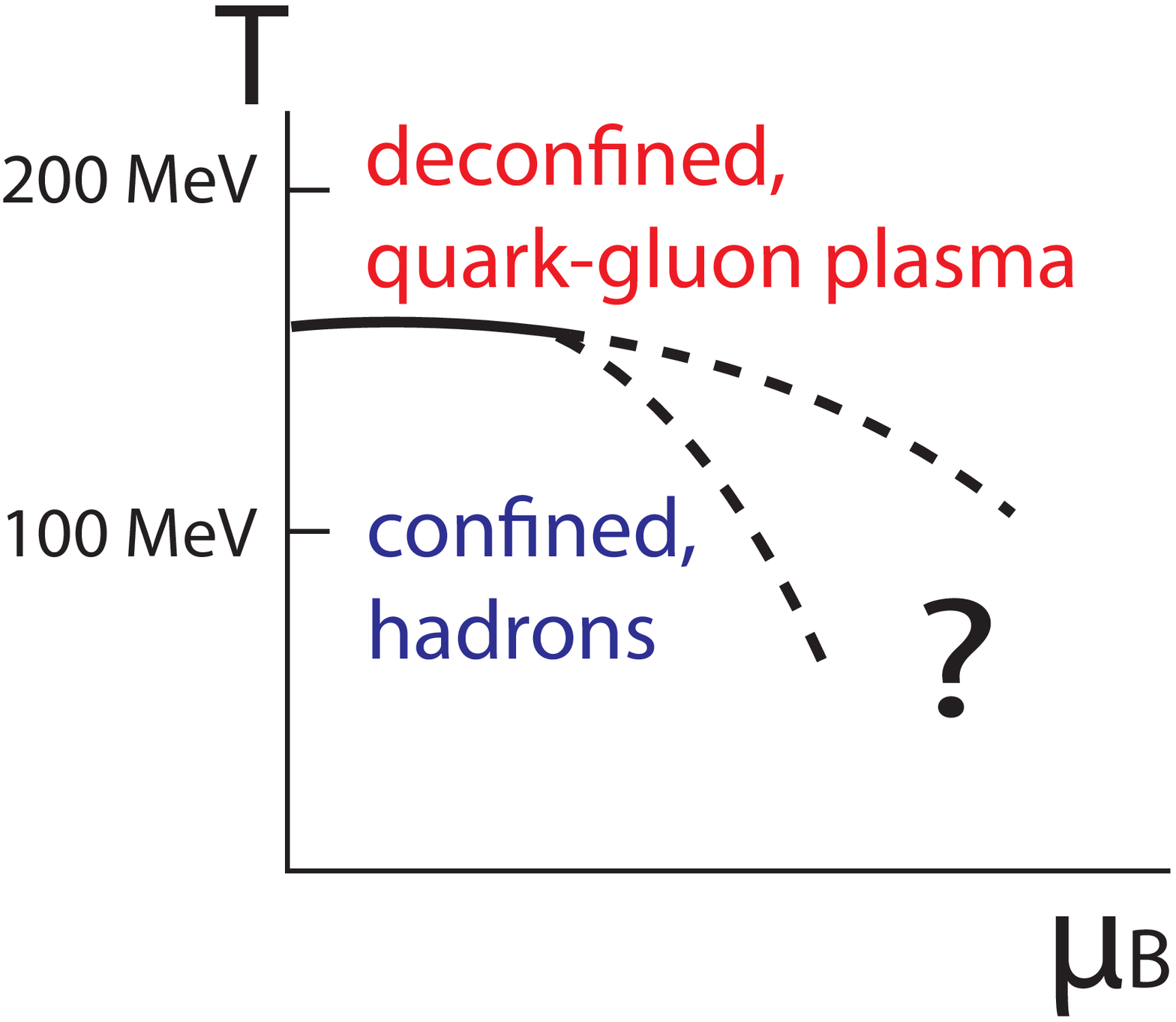}
\hspace*{-6mm}
\includegraphics[angle=0,width=.625\linewidth]{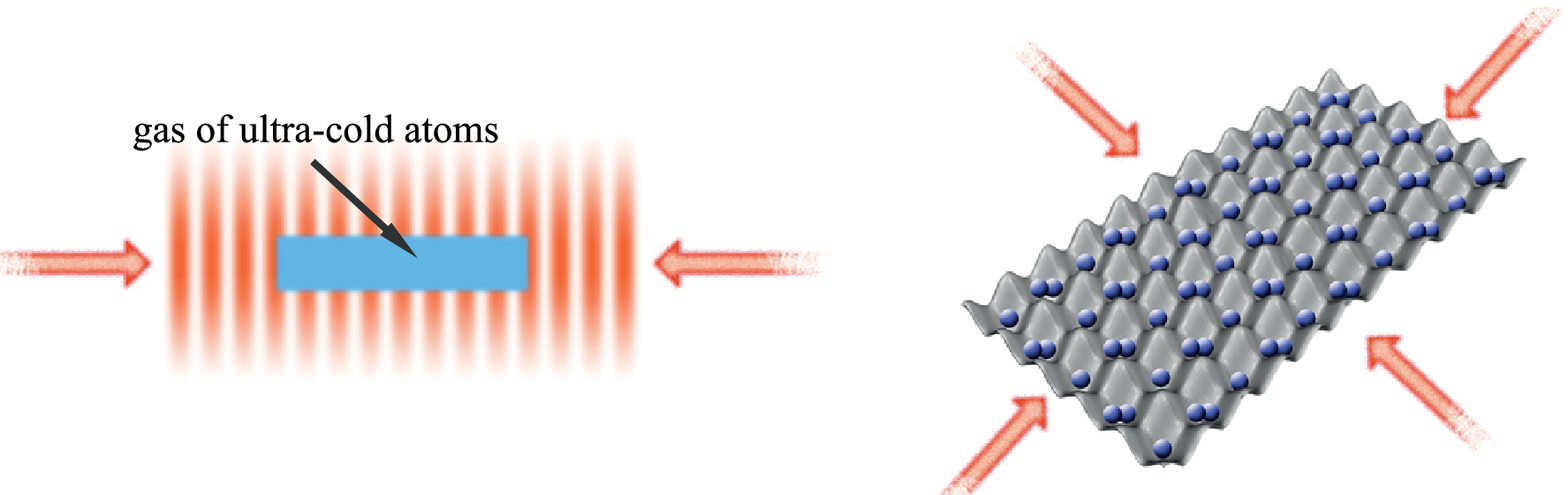}
\caption{On the left: cartoon of the QCD phase diagram,
where an increasing chemical potential $\mu_{B}$ corresponds to
higher baryon density. At low density the crossover temperature
is known, but the phase structure at high density is still
{\it terra incognita.} Center and right: illustrations of an optical 
lattice, built from standing waves of laser light. Ultracold atoms
may be trapped on the sites and thus physically implement a
lattice field theory.}
\label{optlat}
\vspace*{-7mm}
\end{figure}

However, the question what happens at {\em high baryon density} 
is even more difficult to explore, and still open.
There are data from laboratories like the LHC (at CERN) and
RHIC (at BNL), and observations from extremely dense objects, 
in particular neutron stars, as well
as numerous theoretical conjectures. Still, the QCD phase
diagram at high density is one of the major mysteries
within the Standard Model that still persists.
Fig.\ \ref{optlat} (left) shows a cartoon of the unknown 
phase diagram. A large chemical potential $\mu_{B}$ corresponds
to high baryon density\footnote{Intuitively, the chemical potential
can be viewed as the energy, which is required for adding one more 
particle (a baryon in the case of $\mu_{B}$). Technically it amounts
to adding an imaginary part to the momentum component $p_{4}$,
which breaks $\gamma_{5}$-Hermiticity of the lattice Dirac operator.}, 
which occurs under exotic circumstances (the nucleon mass can be 
taken as a reference scale).

The reason why --- unlike other issues --- this hasn't been 
settled yet by lattice simulations is the {\em sign problem:}
the inclusion of $\mu_{B}$ attaches an imaginary part to
the Euclidean action $S_{\rm QCD}$.\cite{PdF}
Thus one faces the problem mentioned in footnote \ref{signprob}; 
the quantity of eq.\ (\ref{prob}) becomes {\em complex,}
\be
p [U] = \left. 
\frac{1}{Z} \exp (- S_{\rm QCD}[U])\right\vert_{\mu_{B} >0} \notin \R \ .
\ee
Obviously, in this setting $p [U]$ does not define any probability, 
and the straightforward approach to simulate QCD as a
statistical system fails.

One could still perform simulations, say with $| \exp (-S_{\rm QCD})|$ 
and incorporate the complex phase {\it a posteriori} by reweighting. 
This is correct in principal, but in practice it
leads to lots of cancellations, and the wanted signal is hard to
extract: for stable statistical errors, the required amount of
data {\em grows exponentially in the volume.} This is the
technical meaning of the notorious ``sign problem'', which usually
prevents us from arriving at conclusive results (except
at small $\mu_{B}$).

Many attempts have been -- and are being --- made to overcome
this problem (Taylor expansion in $\mu_{B}/T$, extrapolation from
imaginary $\mu_{B}$, world line formalism, complex Langevin algorithm
etc.)\cite{PdF}, but there is no real breakthrough so far.
An approach, which bears the potential of a solution, is 
the application of {\em quantum} (rather than classical) {\em computing,}
where the complex phase is naturally incorporated. Digital quantum
computers are still very far from being powerful enough for such 
tasks, but there is rapid progress in analog quantum computing, 
for instance employing ultracold atoms (at nK temperatures)
trapped in optical lattices.\cite{LSA} 
Such lattices are built by the nodes of standing 
laser waves, see Fig.\ \ref{optlat} (center/right).

The corresponding literature is tremendous; for lack of space 
and knowledge, we only mention one recent proposal to quantum
simulate the 2d CP(2) model (or higher CP$(N-1)$ models).\cite{cp2}
This toy model shares a number of qualitative features with QCD: 
asymptotic freedom, a dynamically generated mass gap, 
topological sectors, a unitary global symmetry (which may 
break spontaneously) and even a local symmetry.\cite{cpn1} 

The proposed scenario uses ultracold alkaline earth atoms located 
on the sites of a rectangular optical $L \times L'$ lattice, with 
$L \gg L'$ (Fig.\ \ref{optlat}, center/right).
The relevant degrees of freedom are the nuclear spins, which 
represent an SU(3) field (a staggered atomic site occupation yields 
anti-ferromagnetic coupling). When the (2+1)-d system approaches
its continuum limit (where the correlation length diverges),
it undergoes dimensional reduction\footnote{This is a generic pattern
of D-theory.\cite{Dtheory}.}
(the $L'$-direction becomes negligible), as well as spontaneous 
symmetry breaking SU(3) $\to $ U(2). The low energy effective theory 
for the four NGBs (cf.\ Subsection 3.3) just matches the 2d CP(2) model.

The experimental realization is feasible by means of
established techniques.\cite{cp2} It would allow for the
measurement of quantities like the phase diagram at high
density and real-time dynamics, which are unaccessible to
simulations on classical computers, because of the sign problem. 
This would mean a step forward towards the
long-term goal of a quantum simulation of high density
QCD, which could explore the unknown main part, the 
{\it terra incognita,} of the QCD phase diagram.


\begin{thebibliography}{10}

\bibitem{books} M.\ Creutz, 
{\it Quarks, Gluons and Lattices}
(Cambridge University Press, 1983).
H.J.\ Rothe, 
{\it Lattice Gauge Theories: An Introduction}
(World Scientific, 1992) %Singapore, 1992). 
I.\ Montvay and G.\ M\"{u}nster, 
{\it Quantum Fields on a Lattice} 
(Cambridge University Press, 1994). 
J.\ Smit, {\it Introduction to Quantum Fields on a Lattice}
(Cambridge University Press, 2002).
T.\ DeGrand and C.\ DeTar,
{\it Lattice Methods for Quantum Chromodynamics}
(World Scientific, 2006).
C.\ Gattringer and C.B.\ Lang,
{\it Quantum Chromodynamics on the Lattice}
(Springer, 2009).

\bibitem{WilKog} K.G.\ Wilson and J.B.\ Kogut,
% The Renormalization group and the epsilon expansion
{\it Phys.\ Rept.}\ {\bf 12} (1974) 75.

\bibitem{numrep} W.H.\ Press, S.A.\ Teukolsky, W.T.\ Vetterling 
and B.P. Flannery, {\it Numerical Recipes, The Art of Scientific 
Computing} (Cambridge University Press, 2007 (3rd edition)).

\bibitem{CreutzFreedman} M.\ Creutz and B.\ Freedman,
% A Statistical Approach to Quantum Mechanics {\it Annals of Physics}
{\it Ann.\ Phys.\ (N.Y.)} {\bf 132} (1981) 427.

\bibitem{clusteralgo} R.H.\ Swendsen and J.-S.\ Wang,
% Nonuniversal critical dynamics in Monte Carlo simulations,
{\it Phys.\ Rev.\ Lett.}\ {\bf 58} (1987) 86.
U.\ Wolff,
% Collective Monte Carlo Updating for Spin Systems, 
{\it Phys.\ Rev.\ Lett.}\ {\bf 62} (1989) 361.

\bibitem{clusterfermi} S.\ Chandrasekharan and U.-J.\ Wiese,
% Meron cluster solution of a fermion sign problem
{\em Phys.\ Rev.\ Lett.}\ {\bf 83} (1999) 3116.
% cond-mat/9902128 

\bibitem{HMC} S.\ Duane, A.D.\ Kennedy, B.J.\ Pendleton and D.\ Roweth, 
% Hybrid Monte Carlo,"
{\it Phys.\ Lett.\ B} {\bf 195} (1987) 216.

\bibitem{ParisiWu} G.\ Parisi and Y.-S.\ Wu, {\em Sci.\ Sin.}\
{\bf 24} (1981) 483.

\bibitem{Wilfer} K.G.\ Wilson, in {\it New Phenomena in Subnuclear 
Physics,} ed.\ A.\ Zichichi (Plenum, 1979), p.\ 69.

\bibitem{KoSu} J.B.\ Kogut and L.\ Susskind, 
{\it Phys.\ Rev.}\ D {\bf 11} (1975) 395.

\bibitem{ML} M.\ L\"{u}scher, 
%\emph{Exact chiral symmetry on the lattice and the Ginsparg-Wilson relation}, 
\emph{Phys.\ Lett.}\ B {\bf 428} (1998) 342.

\bibitem{ML2} M.\ L\"{u}scher,
% Abelian chiral gauge theories on the lattice with exact gauge invariance
{\it Nucl.\ Phys.}\ B {\bf 549} (1999) 295.
% hep-lat/9811032

\bibitem{GiWi} P.H.\ Ginsparg and K.G.\ Wilson, 
{\it Phys. Rev.}\ D {\bf 25} (1982) 2649.

\bibitem{Kaplan} D.B.\ Kaplan, {\em Phys.\ Lett.}\ B {\bf 288} (1992) 342.

\bibitem{perfect} U.-J.\ Wiese, {\em Phys.\ Lett.}\ B {\bf 315}
(1993) 417. W.\ Bietenholz and U.-J.\ Wiese,
% Fixed point actions for lattice fermions.
{\em Nucl.\ Phys.\ B (Proc.\ Suppl.)} {\bf 34} (1994) 516; 
% hep-lat/9311016
{\it Phys.\ Lett.}\ B {\bf 378} (1996) 222;
{\it Nucl.\ Phys.}\ B {\bf 464} (1996) 319.
P.~Hasenfratz, V.~Laliena and F.~Niedermayer, 
%\emph{The index theorem in QCD with a finite cut-off}, 
{\it Phys.\ Lett.}\ B {\bf 427} (1998) 125. % [{\tt hep-lat/9801021}].
P.~Hasenfratz, %\emph{Lattice QCD without tuning, mixing and current
%renormalization}, 
{\it Nucl.\ Phys.}\ B {\bf 525} (1998) 401.

\bibitem{overlap} Y.\ Kikukawa and H.\ Neuberger,
% Overlap in odd dimensions
{\it Nucl.\ Phys.}\ B {\bf 513} (1998) 735.
H.\ Neuberger, {\it Phys.\ Lett.}\ B {\bf 417}
(1998) 141; {\it Phys.\ Lett.}\ B {\bf 427} (1998) 353.

\bibitem{revGWR} S.\ Chandrasekharan and U.-J.\ Wiese,
% An Introduction to chiral symmetry on the lattice
{\em Prog.\ Part.\ Nucl.\ Phys.}\ {\bf 53} (2004) 373.
%e-Print: hep-lat/0405024
W.\ Bietenholz,
% Optimized Dirac Operators on the Lattice: 
% Construction, Properties and Applications
{\em Fortsch.\ Phys.}\ {\bf 56} (2008) 107;
%hep-lat/0611030
% W.\ Bietenholz,
% Chiral Fermions on the Lattice
{\it AIP Conf.\ Proc.}\ {\bf 1361} (2011) 245.
% arXiv:1007.0285 [hep-lat] 

\bibitem{PS} M.E.\ Peskin and D.V.\ Schroeder,
{\it An Introduction to Quantum Field Theory} (Westview Press, 1995).

\bibitem{Rainer} For a review, see R.\ Sommer,
% Scale setting in lattice QCD
{\it PoS LATTICE2013} (2014) 015.
%arXiv:1401.3270 [hep-lat]

\bibitem{Marc} M.\ Wagner, S.\ Diehl, T.\ Kuske and J.\ Weber,	
% An introduction to lattice hadron spectroscopy for 
% students without quantum field theoretical background
arXiv:1310.1760 [hep-lat].

\bibitem{prospects} N.\ Brambilla {\it et al.},
% QCD and strongly coupled gauge theories: challenges and perspectives
{\em Eur.\ Phys.\ J.}\ C {\bf 74} (2014) 2981.
% arXiv:1404.3723

\bibitem{Anna} See {\it e.g.}\ Z.\ Fodor, K.\ Holland, J.\ Kuti, 
S.\ Mondal, D.\ Nogradi and C.-H.\ Wong,
%The running coupling of the minimal sextet composite Higgs model
{\em JHEP} {\bf 1509} (2015) 039.
% arXiv:1506.06599 [hep-lat]
A.\ Hasenfratz, R.C.\ Brower, C.\ Rebbi, 
E.\ Weinberg and O.\ Witzel,
% Strongly coupled gauge theories: 
% What can lattice calculations teach us?
arXiv:1510.04635 [hep-lat].

\bibitem{Dark} For recent reviews, see 
R.\ Lewis,
% Dark matter on the lattice
{\em AIP Conf.\ Proc.}\ {\bf 1701} (2016) 090005.
% arXiv:1411.7396 [hep-lat]
G.D.\ Kribs and E.T.\ Neil,
% Review of strongly-coupled composite dark matter
% models and lattice simulations
arXiv:1604.04627 [hep-ph].
% Invited review for IJMPA special issue 
% "Lattice gauge theories beyond QCD"

\bibitem{Alessandra} For a review, see A.\ Feo,
% Predictions and recent results in SUSY on the lattice
{\em Mod.\ Phys.\ Lett.}\ A {\bf 19} (2004) 2387.
% hep-lat/0410012

\bibitem{4dNCU1} Regarding simulations of field theories in
a non-commutaitve space, the study of 4d U(1) gauge theory is 
perhaps closest to particle phenomenology, see
W.\ Bietenholz, J.\ Nishimura, Y.\ Susaki and J.\ Volkholz,
% A Non-perturbative study of 4-D U(1) non-commutative gauge theory: 
% The Fate of one-loop instability
{\em JHEP} {\bf 0610} (2006) 042.
% hep-th/0608072 

\bibitem{MLlect} M.\ L\"{u}scher, 
% Lattice QCD: From quark confinement to asymptotic freedom
{\em Annales Henri Poincar\'{e}} {\bf 4} (2003) 197. %S197-S210
% hep-ph/0211220

\bibitem{stringbreak} See {\it e.g.}\ C.W.\ Bernard {\it et al.}, 
%T.A.\ DeGrand, C.E.\ Detar, P.\ Lacock, S.A.\ Gottlieb, 
%U.M.\ Heller, J.\ Hetrick, K.\ Orginos, D.\ Toussaint, R.L.\ Sugar,
% Zero temperature string breaking in lattice quantum chromodynamics
{\em Phys.\ Rev.}\ D {\bf 64} (2001) 074509.
% hep-lat/0103012

\bibitem{CPPACS} S.\ Aoki {\it et al.} (CP-PACS Collaboration),
% Light hadron spectrum and quark masses from quenched lattice QCD
{\em Phys.\ Rev.}\ D {\bf 67} (2003) 034503.
% hep-lat/0206009

\bibitem{nucstrange} G.S.\ Bali {\it et al.} (QCDSF Collaboration),
% The strange and light quark contributions to the nucleon 
% mass from Lattice QCD
{\em Phys.\ Rev.}\ D {\bf 85} (2012) 054502;
% arXiv:1111.1600 [hep-lat] 
% Strangeness Contribution to the Proton Spin from Lattice QCD 
{\em Phys.\ Rev.\ Lett.}\ {\bf 108} (2012) 222001.
% arXiv:1112.3354 

\bibitem{BMW} S.\ D\"{u}rr {\it et al.} 
(Budapest-Marseille-Wuppertal Collaboration),
% Ab-Initio Determination of Light Hadron Masses
{\it Science} {\bf 322} (2008) 1224.
% arXiv:0906.3599 [hep-lat] 

\bibitem{QCDSFPLB} W.\ Bietenholz {\it et al.} 
(QCDSF-UKQCD Collaboration),
% Tuning the strange quark mass in lattice simulations
{\it Phys.\ Lett.}\ B {\bf 690} (2010) 436.
% arXiv:1003.1114 [hep-lat] 

\bibitem{QCDSFPRD} W.\ Bietenholz {\it et al.} 
(QCDSF-UKQCD Collaboration),
% Flavour blindness and patterns of flavour symmetry 
% breaking in lattice simulations of up, down and strange quarks
{\it Phys.\ Rev.}\ D {\bf 84} (2011) 054509.
% arXiv:1102.5300 [hep-lat] 

\bibitem{FodHoel} Z.\ Fodor and C.\ Hoelbling,
% Light Hadron Masses from Lattice QCD
{\it Rev.\ Mod.\ Phys.}\ {\bf 84} (2012) 449.
% arXiv:1203.4789 [hep-lat]

\bibitem{FLAG} S.\ Aoki {\it et al.} (FLAG Working Group),
% Review of lattice results concerning low-energy particle physics
{\it Eur.\ Phys.\ J.}\ C {\bf 74} (2014) 2890.
%arXiv:1310.8555 [hep-lat] 
 
\bibitem{XPT} S.\ Weinberg, \emph{Physica} A {\bf 96} (1979) 327.
J.\ Gasser and H.\ Leutwyler, {\it Ann.\ Phys.\ (N.Y.)} 
{\bf 158} (1984) 142.

\bibitem{preg} J.\ Gasser and H.\ Leutwyler, 
{\it Phys. Lett.}\ B {\bf 184} (1987) 83.

\bibitem{epsreg} J.\ Gasser and H.\ Leutwyler,
{\it Phys.\ Lett.}\ B {\bf 188} (1987) 477.

\bibitem{deltareg} H.\ Leutwyler, {\em Phys.\ Lett.}\ B {\bf 189}
(1987) 197.

\bibitem{LeuSmi} H.\ Leutwyler and A.V.\ Smilga,
% Spectrum of Dirac operator and role of winding number in QCD
{\em Phys.\ Rev.}\ D {\bf 46} (1992) 5607.

\bibitem{Poul} P.H.\ Damgaard and S.M.\ Nishigaki,
% Universal spectral correlators and massive Dirac operators,
{\em Nucl.\ Phys.}\ B {\bf 518} (1998) 495;
% hep-th/9711023
% Distribution of the k-th smallest Dirac operator eigenvalue
{\em Phys.\ Rev.}\ D {\bf 63} (2001) 045012.
%hep-th/0006111.

\bibitem{RMT} W.~Bietenholz, K.~Jansen and S.~Shcheredin, 
%\emph{Spectral properties of the overlap Dirac operator in QCD}, 
\emph{JHEP} {\bf 0307} (2003) 033. % [{\tt hep-lat/0306022}].
L.~Giusti, M.~L\"uscher, P.~Weisz and H.~Wittig, 
%\emph{Lattice QCD in the $\epsilon$-regime and random matrix theory}, 
\emph{JHEP} {\bf 0311} (2003) 023.
D.~Galletly {\it et al.}\ (QCDSF-UKQCD Collaboration),
%\emph{Quark spectra and light hadron phenomenology from overlap 
%fermions with improved gauge field action},
\emph{Nucl.\ Phys.\ B (Proc.\ Suppl.)} {\bf 129\&130} (2004) 453.
%[{\tt hep-lat/0310028}].

\bibitem{Stani} W.\ Bietenholz and S.\ Shcheredin,
{\em Nucl.\ Phys.}\ B {\bf 754} (2006) 17.
% [hep-lat/0605013].

\bibitem{AA} P.H.\ Damgaard, P. Hern\'{a}ndez, K.\ Jansen, 
M.\ Laine and L. Lellouch,
% Finite size scaling of vector and axial current correlators
{\em  Nucl.\ Phys.}\ B {\bf 656} (2003) 226.
% hep-lat/0211020
W.~Bietenholz, T.~Chiarappa, K.~Jansen, K.-I.~Nagai
and S.~Shcheredin,
%\emph{Axial Correlation Functions in the epsilon-Regime: 
%a Numerical Study with Overlap Fermions},
\emph{JHEP} {\bf 0402} (2004) 023. % [{\tt hep-lat/0311012}].
H.\ Fukaya, S.\ Hashimoto and K.\ Ogawa,
{\it Prog.\ Theor.\ Phys.} {\bf 114} (2005) 451.

\bibitem{0modes} L.\ Giusti, P.\ Hern\'{a}ndez, M.\ Laine, P.\ Weisz
and H.\ Wittig,
% Low-energy couplings of QCD from topological zero mode wave functions
{\em JHEP} {\bf 0401} (2004) 003.
% hep-lat/0312012

\bibitem{Giusti} L.\ Giusti,
% Recent Progress on Chiral Symmetry Breaking in QCD
{\it PoS LATTICE2015} (2015) 001.
% arXiv:1511.08786 [hep-lat] 

\bibitem{Tibur} M.E.\ Matzelle and B.C.\ Tiburzi,
% Low-Energy QCD in the Delta Regime
{\em Phys.\ Rev.}\ D {\bf 93} (2016) 034506.
% arXiv:1512.05286 [hep-lat] 

\bibitem{HasNie} P.\ Hasenfratz and F.\ Niedermayer,
% Finite size and temperature effects in the AF Heisenberg model
{\it Z.\ Phys.}\ B {\bf 92} (1993) 91. % hep-lat/9212022

\bibitem{Has10} P.\ Hasenfratz,
%The QCD rotator in the chiral limit.
{\em Nucl.\ Phys.}\ B {\bf 828} (2010) 201.

\bibitem{NieWei} F.\ Niedermayer and P.\ Weisz,
% Matching effective chiral Lagrangians with dimensional 
% and lattice regularization
arXiv:1601.00614 [hep-lat].

\bibitem{QCDSFdelta} W.\ Bietenholz {\it et al.} (QCDSF Collaboration), 
% M.\ G\"{o}ckeler, R.\ Horsley, Y.\ Nakamura, D.\ Pleiter, P.E.L.\ Rakow, 
% G.\ Schierholz and J.M.\ Zanotti (QCDSF Collaboration), 
{\it Phys.\ Lett.}\ B {\bf 687} (2010) 410;
% arXiv:1002.1696 [hep-lat] 
% QCD in the delta-Regime, with N.\ Cundy
{\em J.\ Phys.\ Conf.\ Ser.}\ {\bf 287} (2011) 012016.
% arXiv:1103.3311 [hep-lat]

\bibitem{ILDG} http://plone.jldg.org/wiki/index.php/Main$_{-}$Page

\bibitem{Borsanyi15} Sz.\ Bors\'{a}nyi {\it et al.},
% Ab initio calculation of the neutron-proton mass difference
{\it Science} {\bf 347} (2015) 1452.
%arXiv:1406.4088 [hep-lat]  uncertainty 0.3 MeV

\bibitem{munonzero} T.\ Blum {\it et al.},
% Electromagnetic mass splittings of the low lying hadrons 
% and quark masses from 2+1 flavor lattice QCD+QED
{\em Phys.\ Rev.}\ D {\bf 82} (2010) 094508.
% arXiv:1006.1311 [hep-lat] 
R.\ Horsley {\it et al.} (QCDSF-UKQCD Collaboration),
% QED effects in the pseudoscalar meson sector
{\em JHEP} {\bf 1604} (2016) 093.
% arXiv:1509.00799 [hep-lat]
Z.\ Fodor {\it et al.},
% Z. Fodor, C. Hoelbling, S. Krieg, L. Lellouch, Th. Lippert, 
% A. Portelli, A. Sastre, K.K. Szabo, L. Varnhorst
% Up and down quark masses and corrections to Dashen's theorem 
% from lattice QCD and quenched QED
arXiv:1604.07112 [hep-lat].

\bibitem{Leinw} D.\ Leinweber {\it et al.},
% W.\ Kamleh, A.\ Kiratidis, Z.-W.\ Liu,
% S.\ Mahbub, D.\ Roberts, F.\ Stokes, A.W.\ Thomas and J.\ Wu,
% N* Spectroscopy from Lattice QCD: The Roper Explained
arXiv:1511.09146 [hep-lat].

\bibitem{HPQCD} I.F.\ Allison, C.T.H.\ Davies, A.\ Gray, A.S.\ Kronfeld, 
P.B.\ Mackenzie and J.N.\ Simone,
% Mass of the Bc meson in three-flavor lattice QCD
% HPQCD and Fermilab Lattice and UKQCD Collaborations
{\em Phys.\ Rev.\ Lett.}\ {\bf 94} (2005) 172001.
% hep-lat/0411027  Gitter: M_B_c = 6.30(2) GeV

\bibitem{CDF} A.\ Abulencia {\it et al.} (CDF Collaboration),
% Evidence for the exclusive decay B±c→J/ψπ± and measurement of 
% the mass of the Bc meson
{\em Phys.\ Rev.\ Lett.}\ {\bf 96} (2006) 082002.
% hep-ex/0505076  Experiment: M_B_c = 6.286(5) GeV

\bibitem{PDG} K.A.\ Olive {\it et al.} (Particle Data Group), 
{\em Chin.\ Phys.}\ C {\bf 38} (2014) 090001 (and 2015 update).
% PDG 2015: 6275.1(1.0) MeV

\bibitem{Rainer2} L.\ Del Debbio, H.\ Panagopoulos and E.\ Vicari,
% Theta dependence of SU(N) gauge theories
{\em JHEP} {\bf 0208} (2002) 044.
% arXiv:hep-th/0204125
L.\ Del Debbio, G.M.\ Manca and E. Vicari,
% Critical slowing down of topological modes
{\em Phys.\ Lett.}\ B {\bf 594} (2004) 315.
S.\ Schaefer, R.\ Sommer and F.\ Virotta
(ALPHA Collaboration),
% Critical slowing down and error analysis in lattice QCD simulations
{\em Nucl.\ Phys.}\ B {\bf 845} (2011) 93.
% arXiv:1009.5228 [hep-lat]

\bibitem{ovHF} W.\ Bietenholz, {\it Eur.\ Phys.\ J.}\ C {\bf 6}
(1999) 537. W.\ Bietenholz and I.\ Hip,
{\it Nucl.\ Phys.}\ B {\bf 570} (2000) 423.
W.\ Bietenholz, {\it Nucl.\ Phys.}\ B {\bf 644} (2002) 223.

\bibitem{WiVe} E.\ Witten, {\em Nucl.\ Phys.}\ B {\bf 156} (1979) 269. 
G.\ Veneziano, {\em Nucl.\ Phys.}\ B {\bf 159} (1979) 213.

\bibitem{AFHO} S.\ Aoki, H.\ Fukaya, S.\ Hashimoto and T.\ Onogi,
{\it Phys.\ Rev.}\ D {\bf 76} (2007) 054508. % [arXiv:0707.0396 [hep-lat]].

\bibitem{AFHOresults} S.\ Aoki {\it et al.}
(JLQCD and TWQCD Collaborations),
{\it Phys.\ Lett.}\ B {\bf 665} (2008) 294.
 H.\ Fukaya {\it et al.} (JLQCD Collaboration), 
% S.\ Aoki, G.\ Cossu, S.\ Hashimoto, 
% T.\ Kaneko and J.\ Noaki (JLQCD Collaboration),
{\it PoS LATTICE2014} (2014) 323. %[arXiv:1411.1473 [hep-lat]].
I.\ Bautista {\it et al.}, 
% W.\ Bietenholz, A.\ Dromard, U.\ Gerber, 
% C.P.\ Hofmann, H.\ Mej\'{\i}a-D\'{\i}az and M.\ Wagner,
% Measuring the Topological Susceptibility in a Fixed Sector
{\it Phys.\ Rev.}\ D {\bf 92} (2015) 114510. %[arXiv:1503.06853 [hep-lat]].

\bibitem{Arthur16} A.\ Dromard, talk presented at {\em Excited QCD},
Lisbon, March 2016.

\bibitem{slab} W.\ Bietenholz, P.\ de Forcrand and U.\ Gerber,
% Topological Susceptibility from Slabs
{\it JHEP} {\bf 1512} (2015) 070. % [arXiv:1509.06433 [hep-lat]].

\bibitem{LSD} R.C.\ Brower {\it et al.} (LSD Collaboration),	
% Maximum-Likelihood Approach to Topological Charge Fluctuations 
% in Lattice Gauge Theory
{\it Phys.\ Rev.}\ D {\bf 90} (2014) 014503.
% [arXiv:1403.2761 [hep-lat]].

\bibitem{Egri} G.I.\ Egri, Z.\ Fodor, S.D.\ Katz and K.K.\ Szab\'{o},
% Topology with dynamical overlap fermions
{\it JHEP} {\bf 0601} (2006) 049.
% [hep-lat/0510117].

\bibitem{topfreez} H.\ Fukaya {\it et al.},
% Two-flavor lattice QCD simulation in the epsilon-regime 
% with exact chiral symmetry 
{\it Phys.\ Rev.\ Lett.}\ {\bf 98} (2007) 172001; % [hep-lat/0702003];
% Lattice study of meson correlators in the epsilon-regime of two-flavor QCD
{\it Phys.\ Rev.}\ D {\bf 76} (2007) 054503. % [arXiv:0711.4965 [hep-lat]].\\
S.~Aoki {\it et al.} (JLQCD Collaboration),
%``Two-flavor QCD simulation with exact chiral symmetry,''
{\it Phys.\ Rev.}\ D {\bf 78} (2008) 014508. % [arXiv:0803.3197 [hep-lat]].\\
Sz.\ Bors\'{a}nyi {\it et al.}, 
%Z.\ Fodor, S.D.\ Katz, S.\ Krieg, T.\ Lippert, D.\ Nogradi, 
%F.\ Pittler, K.K.\ Szabo and B.C. Toth,
% QCD thermodynamics with continuum extrapolated dynamical overlap fermions
arXiv:1510.03376 [hep-lat]. 

\bibitem{Liao} A.\ Laio, G.\ Martinelli and F.\ Sanfilippo,
% Metadynamics Surfing on Topology Barriers: the CP(N-1) Case
arXiv:1508.07270 [hep-lat].

\bibitem{openbc} M.\ L\"{u}scher,
%Properties and uses of the Wilson flow in lattice QCD
{\it JHEP} {\bf 1008} (2010) 071.
%[arXiv:1006.4518 [hep-lat]]
M.\ L\"{u}scher and S.\ Schaefer,
% Lattice QCD without topology barriers
{\it JHEP} {\bf 1107} (2011) 036.
% [arXiv:1105.4749 [hep-lat]].
G.\ McGlynn and R.D.\ Mawhinney,
% Diffusion of topological charge in lattice QCD simulations
{\em Phys.\ Rev.}\ D {\bf 90} (2014) 074502.
% arXiv:1406.4551 [hep-lat] 

\bibitem{Mages} S.\ Mages, B.C.\ T\'{o}th, Sz.\ Bors\'{a}nyi, Z.\ Fodor, 
S.\ Katz and K.K.\ Szab\'{o},
%Lattice QCD on Non-Orientable Manifolds
arXiv:1512.06804 [hep-lat].

\bibitem{BCNW} R.\ Brower, S.\ Chandrasekharan, J.W.\ Negele
and U.-J.\ Wiese, 
% QCD at fixed topology 
{\it Phys.\ Lett.}\ B {\bf 560} (2003) 64. % [hep-lat/0302005].

\bibitem{Frankfurt} A.\ Dromard and M.\ Wagner,
% Extracting hadron masses from fixed topology simulations 
{\it Phys.\ Rev.}\ D {\bf 90} (2014) 074505.
%[arXiv:1404.0247 [hep-lat]].

\bibitem{actpol} A.\ Dromard, W.\ Bietenholz, U.\ Gerber, 
H.\ Mej\'{\i}a-D\'{\i}az and M.\ Wagner,
% Simulations at fixed topology: fixed topology versus 
% ordinary finite volume corrections
{\it Acta Phys.\ Polon.\ Supp.}\ {\bf 8} (2015) 2, 391;
%[arXiv:1505.03435 [hep-lat]];
% Combining ordinary and topological finite volume effects 
% for fixed topology simulations
arXiv:1510.08809 [hep-lat].

\bibitem{BCNWtest} W.\ Bietenholz, C.\ Czaban, A.\ Dromard, 
U.\ Gerber, C.P.\ Hofmann, H.\ Mej\'{\i}a-D\'{\i}az and M.\ Wagner,
% Interpreting Numerical Measurements in Fixed Topological Sectors
arXiv:1603.05630 [hep-lat].

\bibitem{Ludmilla} For reviews, see {\it e.g.}\ L.\ Levkova,
% QCD at nonzero temperature and density
{\em PoS LATTICE2011} (2011) 011.
% arXiv:1201.1516 [hep-lat]
P.\ de Forcrand, O.\ Philipsen and W.\ Unger,
% QCD phase diagram from the lattice at strong coupling
{\em PoS CPOD2014} (2015) 073.
% arXiv:1503.08140 [hep-lat] 
H.-T.\ Ding, F.\ Karsch and S.\ Mukherjee,
% Thermodynamics of strong-interaction matter from Lattice QCD
{\em Int.\ J.\ Mod.\ Phys.}\ E {\bf 24} (2015) 1530007.
% arXiv:1504.05274 [hep-lat]

\bibitem{Tcross} Sz.\ Bors\'{a}nyi {\it et al.}
(Wuppertal-Budapest Collaboration),
% Is there still any T_c mystery in lattice QCD? 
% Results with physical masses in the continuum limit III
{\em JHEP} {\bf 1009} (2010) 073.
% arXiv:1005.3508 [hep-lat]
E.\ Laermann,
% Recent results from high temperature lattice QCD
{\em Phys.\ Part.\ Nucl.}\ {\bf 46} (2015) 740.

\bibitem{PdF} For a review, see P.\ de Forcrand,
% Simulating QCD at finite density
{\em PoS LATTICE2009} (2009) 010.
% [arXiv:1005.0539 [hep-lat]].

\bibitem{LSA} M.\ Lewenstein, A.\ Sanera and V.\ Ahufinger,
{\it Ultracold Atoms in Optical Lattices}
(Oxford University Press, 2012).
U.-J.\ Wiese,
% Ultracold Quantum Gases and Lattice Systems: Quantum Simulation of Lattice Gauge Theories
{\it Annalen Phys.}\ {\bf 525} (2013) 777.
%arXiv:1305.1602 [quant-ph] 
E.\ Zohar, J.I.\ Cirac and B.\ Reznik,
% Quantum Simulations of Lattice Gauge Theories using 
%Ultracold Atoms in Optical Lattices
{\it Rep.\ Prog.\ Phys.}\ {\bf 79} (2016) 014401.
%arXiv:1503.02312 [quant-ph]

\bibitem{cp2} C.\ Laflamme {\it et al.},
%W.\ Evans, M.\ Dalmonte, U.\ Gerber, H.\ Mej\'{\i}a-D\'{\i}az, 
%W.\ Bietenholz, U.-J.\ Wiese and P.\ Zoller, 
{\it Ann.\ Phys.\ (N.Y.)} {\bf 370} (2016) 117;
%arXiv:1507.06788 [quant-ph];
arXiv:1510.08492 [hep-lat].

\bibitem{cpn1} A.\ D'Adda, M.\ L\"{u}scher and P.\ Di Vecchia, 
{\it Nucl.\ Phys.}\ B {\bf 146} (1978) 63.

\bibitem{Dtheory} S.\ Chandrasekharan and U.-J.\ Wiese,
% Quantum link models: A Discrete approach to gauge theories
{\em Nucl.\ Phys.}\ B {\bf 492} (1997) 455.

\end{thebibliography}
\end{document}